\documentclass[twocolumn,epjc3]{svjour3}

\usepackage[utf8]{inputenc}
\usepackage[T1]{fontenc}

\usepackage[numbers,sort&compress]{natbib}

\usepackage{graphicx}
\usepackage{scalefnt}

\usepackage{amsmath}
\usepackage{amssymb}
\usepackage{amsfonts}
\usepackage{mathtools}
\usepackage{braket}

\usepackage{txfonts}
\usepackage{microtype}

\usepackage[dvipsnames]{xcolor}

\usepackage{enumitem}
\newlist{renumerate}{enumerate}{1}
\setlist[renumerate]{label=(\roman*),leftmargin=*}
\newlist{aenumerate}{enumerate}{1}
\setlist[aenumerate]{label=(\arabic*),leftmargin=*}

\newcommand{\subalign}[1]{
\begin{subequations}
\begin{align}
#1
\end{align}
\end{subequations}
}

\usepackage{epstopdf}
\newcommand{\includegrtex}[2][]{\includegraphics[#1]{#2.tex}}

\usepackage{listings}
\lstdefinelanguage{amc}{
    keywords={declare,true,false,sum},
    sensitive=true,
    comment=[l]{\#},
    string=[b]{"}
}
\lstnewenvironment{amclisting}{\lstset{language=amc,mathescape=true}}{}

\usepackage{hyperref}

\newcommand{\clebsch}[6]{\begingroup\setlength{\arraycolsep}{0.1em}\left(\hskip -\arraycolsep\begin{array}{cc|c} #1 & #3 & #5 \\ #2 & #4 & #6 \end{array}\hskip -\arraycolsep\right)\endgroup}
\newcommand{\threej}[6]{\begingroup\setlength{\arraycolsep}{0.1em}\left(\hskip -\arraycolsep\begin{array}{ccc} #1 & #3 & #5 \\ #2 & #4 & #6 \end{array}\hskip -\arraycolsep\right)\endgroup}
\newcommand{\trij}[3]{\begingroup\setlength{\arraycolsep}{0.2em}\begin{Bmatrix} #1 & #2 & #3  \end{Bmatrix}\endgroup}
\newcommand{\sixj}[6]{\begingroup\setlength{\arraycolsep}{0.2em}\begin{Bmatrix} #1 & #2 & #3 \\ #4 & #5 & #6 \end{Bmatrix}\endgroup}

\newcommand{\ninej}[9]{\begingroup\setlength{\arraycolsep}{0.2em}\begin{Bmatrix} #1 & #2 & #3 \\ #4 & #5 & #6 \\ #7 & #8 & #9 \end{Bmatrix}\endgroup}

\newcommand{\diagram}[1]{\begin{center}\includegrtex{#1}\end{center}}

\newcommand{\prog}[2]{
\begin{description}[font=\normalfont\sffamily,style=nextline]
\item[#1] #2
\end{description}
}

\newcommand{\pr}{\prime}

\newcommand{\jhat}[1]{\ensuremath{\hat{#1} }}

\newcommand{\obs}[1]{ \textcolor{blue}{} }

\newcommand{\Gsym}{\ensuremath{G_\text{sym} }}
\newcommand{\Gham}{\ensuremath{G_\text{Ham} }}
\newcommand{\Gref}{\ensuremath{G_\text{ref} }}
\newcommand{\Gbas}{\ensuremath{G_\text{bas} }}

\newcommand{\op}[1]{\ensuremath{\boldsymbol{#1}}}

\renewcommand{\vec}[1]{\ensuremath{\boldsymbol{#1}}}

\newcommand{\AMC}{\textsf{AMC}}
\newcommand{\amc}{\texttt{amc}}

\journalname{Eur. Phys. J. A}
\renewcommand{\email}[1]{e-mail: \href{mailto:#1}{#1}}

\begin{document}

\allowdisplaybreaks



\title{Symmetry reduction of tensor networks in many-body theory}
\subtitle{I. Automated symbolic evaluation of $SU(2)$ algebra}

\author{A.~Tichai\thanksref{ad:mpik,em:at} \and R. Wirth\thanksref{ad:frib,em:rw} \and J.~Ripoche\thanksref{ad:ceadam,em:jr} \and T.~Duguet\thanksref{ad:irfu,em:td}}
\date{Received: \today{} / Accepted: date}

\thankstext{em:at}{\email{alexander.tichai@physik.tu-darmstadt.de}}
\thankstext{em:rw}{\email{wirth@frib.msu.edu}}
\thankstext{em:jr}{\email{julien.ripoche@cea.fr}}
\thankstext{em:td}{\email{thomas.duguet@cea.fr}}

\institute{%
\label{ad:mpik}%
Max-Planck-Institut f\"ur Kernphysik, Heidelberg, Germany \\
Institut f\"ur Kernphysik, Technische Universit\"at Darmstadt, Darmstadt, Germany \\
ExtreMe Matter Institute EMMI, GSI Helmholtzzentrum f\"ur Schwerionenforschung GmbH, Darmstadt, Germany \\
ESNT, CEA-Saclay, DRF, IRFU, D\'epartement de Physique Nucl\'eaire, Universit\'e de Paris Saclay, 91191 Gif-sur-Yvette, France
\and%
\label{ad:frib}%
Facility for Rare Isotope Beams, Michigan State University, East Lansing, Michigan 48824, USA
\and%
\label{ad:ceadam}%
CEA, DAM, DIF, 91297 Arpajon, France
\and%
\label{ad:irfu}%
IRFU, CEA, Universit\'e Paris-Saclay, 91191 Gif-sur-Yvette, France  \\
KU Leuven, Instituut voor Kern- en Stralingsfysica, 3001 Leuven, Belgium
}

\maketitle


\begin{abstract}
The ongoing progress in (nuclear) many-body theory is accompanied by an ever-rising increase in complexity of the underlying formalisms used to solve the stationary Schr\"odinger equation. The associated working equations at play in state-of-the-art  \textit{ab initio} nuclear many-body methods can be analytically reduced with respect to angular-momentum, i.e. $SU(2)$, quantum numbers whenever they are effectively employed in a symmetry-restricted context. The corresponding procedure constitutes a tedious and error-prone but yet an integral part of the implementation of those many-body frameworks. Indeed, this symmetry reduction is a key step to advance modern simulations to higher accuracy since the use of symmetry-adapted tensors can decrease the computational complexity by orders of magnitude.

While attempts have been made in the past to automate the (anti\nobreakdash-) commutation rules linked to Fermionic and Boson\-ic algebras at play in the derivation of the working equations, there is no systematic account to achieve the same goal for their symmetry reduction. In this work, the first version of an automated tool performing graph-theory-based angular-momentum reduction is presented. Taking the symmetry-unrestricted expressions of a generic tensor network as an input, the code provides their angular-momentum-reduced form in an error-safe way in a matter of seconds. Several state-of-the-art many-body methods serve as examples to demonstrate the generality of the approach and to highlight the potential impact on the many-body community.
\end{abstract}

\section*{PROGRAM SUMMARY}
\begin{description}[font=\normalfont\itshape]
  \item[Program title:] \AMC{}
  \item[Licensing provisions:] GNU General Public License Version 3 or later
  \item[Programming language:] Python 3
  \item[Repository and DOI:] \href{https://github.com/radnut/amc}{\nolinkurl{github.com/radnut/amc}} \\ DOI: \href{https://doi.org/}{\nolinkurl{10.5281/zenodo.3663059}}
  \item[Nature of problem:]
    Numerical implementations of state-of-the-art many-body approaches require extensive use of angular-momentum algebra to derive the spherically reduced form of working equations.
    This derivation takes a lot of effort and is prone to errors.
  \item[Solution method:]
    Angular-momentum objects are simplified via identification of subgraphs in a suitably defined network called \emph{Yutsis graph}.
    With this, a spherical reduction of a tensor network is obtained and all quantities are expressed in terms of their $m$-independent, reduced analogues.
    The reduction process is fully automated, limiting the potential for human error.
  \item[Additional comments:]
    The reduction is formulated as a transformation of abstract syntax trees that facilitates post-processing into different output formats, as well as automated code generation.
\end{description}

\section{Introduction}
\label{sec:intro}

In recent years, \emph{ab initio} nuclear many-body theory has undergone a major renewal. In this process, expansion methods have become prominent in large-scale applications to mid-mass nuclei. The success obtained within the last two decades is leading to the design of more and more advanced approaches to continuously refine the accuracy of the calculations and extend them systematically to an even larger portion of the nuclear chart. This rise in the degree of sophistication of state-of-the-art many-body expansion schemes is leading to an increase of the formal complexity that is now at the edge of what is humanly processable.

When following the \emph{ab initio} philosophy to solve the stationary Schr\"odinger equation, quasi-exact approaches based on Monte Carlo techniques~\cite{Gezerlis2013,Carlson:2015,Lynn2017,Lynn2019} or configuration interaction (CI)~\cite{NaQu09,BaVa13} are limited by their computational scaling to the lightest systems.
Moving to the realm of medium- and heavy-mass nuclei involves the use of expansion many-body techniques building a wave-function parametrization on top of a conveniently chosen reference state. These methods display a polynomial scaling with system size, the degree of the polynomial increasing with the targeted accuracy, i.e., with the order at which the expansion is truncated.
This computational advantage typically comes at the price of being restricted to working in a non-variational scheme. Examples of such approaches are many-body perturbation theory (MBPT)~\cite{Go57,Shavitt2009,Tichai2016,Hu:2016txm,Xu17,Tichai:2018mll,Tichai:2018ncsmpt,Hu18arxiv,Tichai:2020dna,Demol2020a}, coupled-cluster (CC) theory~\cite{KoDe04,Binder2013,Ja14,Si15,Duguet:2014jja,Duguet:2015yle,Qiu:2019edx}, self-consistent Green's function (SCGF) theory~\cite{Dickhoff:2004xx,CiBa13,Carbone:2013eqa,Soma:2011aj,SoCi13} or the in-medium similarity renormalization group (IMSRG) method~\cite{Tsukiyama:2011,Tsukiyama:2012,Hergert2013,Bo14,H15,Morris:2017vxi,Stroberg2017,Parzuchowski2017,Hergert:2018wmx}, all of which provide a consistent description of (at least) ground-state observables in nuclear many-body systems. In quantum chemistry in particular, MBPT and CC theories have a long tradition and both frameworks have been derived and implemented at very high truncation orders~\cite{Shavitt2009}. Although every member of the aforementioned approaches can be applied to much higher masses and larger system sizes than exact methods, the truncation levels needed for high-accuracy calculations require substantial effort in the derivation of the formalisms and for their numerical implementations.

While in earlier works the working equations were derived by hand, the rising computational power and the development of computer-aided algebraic manipulation tools have facilitated the derivation of more advanced truncation schemes in modern many-body approaches, many of which have undergone their pioneering studies in quantum chemistry~\cite{Paldus1973,Ka76,Csepes1988,Lyons:1994ew,Xiao2013}. A shining example is the tensor contraction engine that was developed in close collaboration with computer scientists and has been one of the most powerful tools to extend quantum-chemistry calculations to higher accuracy by generating working equations and source code for large-scale distributed implementations~\cite{Hi03}.

Even though large progress has been made in the development of software supporting the formal developments at play in quantum many-body research, only few are directly dedicated to the nuclear many-body problem~\cite{Arthuis:2018yoo}. While sharing many formal similarities with its electronic counterpart the nuclear many-body problem differs in two key points requiring a dedicated attention
\begin{aenumerate}
\item At the mean-field level, single-nucleon states carry good total angular momentum $\vec{j}=\vec{l}+\vec{s}$, i.e., one must employ the so-called $j$-coupling scheme to define appropriate one-nucleon states. Contrarily, electrons carry a well-defined projection $s_z$ of the intrinsic spin and are, thus, best described on the basis of the so-called $ls$-coupling. The main consequence is that nucleons orbit in energy shells characterized by a greater degree of degeneracy, thus leading to the large dominance of open-shell ground-states, i.e., degenerate systems, over the nuclear chart.
\item The inclusion of three-body forces in a realistic nuclear Hamiltonian is mandatory to ensure a quantitatively correct description of nuclear observables, i.e. one is bound to use
\begin{align}
H = T + V + W + ... \, ,
\end{align}
where $T$ is the kinetic energy operator whereas $V$ and $W$ are two- and three-body potentials, respectively.
\end{aenumerate}
While expansion many-body methods are first formulated in terms of a generic single-particle basis, their actual implementations typically exploit symmetry properties of the basis functions and of the targeted many-body state, e.g., with respect to angular-momentum or parity quantum numbers. The adaptation of the generic formalism to a specific symmetry group defines a \emph{symmetry reduction} of the many-body formalism. The goal is to use reduced many-body tensors associated to irreducible representations (IRREPs) of the symmetry group to pre-process a subset of the summations at play in the tensor networks defining the working equations.

A simple, yet representative, example is the pre-process\-ing of spin summations in spin-restricted quantum chemistry calculations. The counterpart in nuclear structure theory relates to the exploitation of \emph{rotational invariance} associated with the conservation of \emph{total} angular momentum and encoded in terms of the $SU(2)$ nonabelian Lie group. In this particular case the reduction scheme will be referred as the \emph{angular-momentum reduction (AMR)}.

Eventually, it turns out that the AMR poses a nontrivial problem  requiring the same amount of effort that the derivation of the initial working equations. However, there exists a highly systematic and elegant way to deal with this task that is close in spirit to the use of Feynman's diagrams as a mnemonic device to represent physical processes.

Consequently, it is highly desirable to parallel the efforts done to automatize the generation of working equations by devising a framework that automatically performs the tedious symmetry reduction in an error-safe way. Currently, there is -- to the best of our knowledge -- no open-source library that can deal with the requirements imposed by nuclear structure many-body methods to perform symbolic manipulations of angular-momentum algebra. Typically, existing software is restricted to the numerical evaluation of coupling coefficients instead of performing symbolic manipulations including the simplification of complex tensor networks. There have been similar attempts for symbolic simplifications of angular-momentum expressions before without formally connecting it to many-body theory~\cite{Wormer2006}.

Therefore, the goal of the source code accompanying the present document is to support the implementation of advanced many-body frameworks in nuclear structure in an error-free way. Of course, this does not resolve the problem of writing an efficient and error-free numerical implementation of the symmetry-reduced formalism itself. While the generation of the source code is envisioned, it is, however, beyond the scope of the present work.

The document is organized as follows.  In Sec.~\ref{sec:sym} the notion of symmetry in the context of many-body theory is introduced using a group theory formulation. Section~\ref{sec:ama} focuses on the angular-momentum algebra and its relation to states and operators. In Sec.~\ref{sec:amcdiag} the diagrammatic allowing for the handling of the $SU(2)$ algebra is laid out and the simplification rules for the graph theory reformulation of tensor networks are presented. Section~\ref{sec:appli} discusses several state-of-the-art many-body approaches that serve as pedagogical examples to demonstrate the generality of the approach. Ultimately, an outlook is provided in Sec.~\ref{sec:conc}.

\section{Symmetries and many-body theory}
\label{sec:sym}

\subsection{Symmetry group}

Physical symmetries impact many-body formalisms at various stages of their elaboration. The existence of symmetries in finite systems is intimately connected to conservation laws, e.g., the existence of $U(1)$ global gauge symmetry corresponds to particle-number conservation while $SU(2)$ symmetry corresponds to angular momentum conservation. Mathematically, the \emph{invariance} of a quantum system, characterized by its Hamiltonian $H$, is encoded in terms of transformation properties imposed by a \emph{symmetry group} $G_\text{Ham}$ whose action leaves the physical system invariant or, equivalently, the existence of a unitary linear representation $U$ acting on the space of states such that
\begin{align}
H = U(g) H U^\dagger(g)  \quad (\forall g \in G) \, , \label{conservation1}
\end{align}
which can be rewritten as
\begin{align}
[ H , U(g) ] =0 \quad (\forall g \in G) \, . \label{conservation2}
\end{align}
Given the eigenstates of the Hamiltonian
\begin{align}
H \ket{ \Psi_k } = E_k \ket{ \Psi_k } ,
\end{align}
Eqs.~(\ref{conservation1}-\ref{conservation2}) stipulate that the transformed states
\begin{align}
\ket{\Psi_k (g) } \equiv U(g) \ket{\Psi_k }  \quad (\forall g \in G) \, ,
\end{align}
are also eigenstates with the same eigenvalues.

In the case of discrete symmetries such as parity or time reversal the corresponding symmetry group is finite, e.g. $\mathbb{Z}_2$. Contrarily, continuous symmetries correspond to Lie groups allowing for a continuous parametrization of the (infinite number of) group elements in terms of a finite set of parameters. The present focus is on the nonabelian $SU(2)$ Lie group associated with rotational invariance of nuclear systems. Relevant details about this symmetry group are provided in Sec.~\ref{su2}.

Eventually, symmetries enter the formulation of (nuclear) quantum many-body methods at three different levels
\begin{aenumerate}
\item the \emph{symmetry group of the Hamiltonian} $G_\text{Ham}$ specifying the invariance of the physical system under a given set of transformations along with the symmetry quantum numbers carried by its many-body eigenstates,
\item the \emph{symmetry group of the single-particle basis} $G_\text{bas}$ specifying the symmetry properties of the computational basis,
\item the \emph{symmetry group of the reference state} $G_\text{ref}$ employed in expansion methods specifying the symmetries of the auxiliary many-body problem that is solved to construct the reference state.
\end{aenumerate}
While the symmetry group of the Hamiltonian is fixed by the physical system under consideration, the symmetry properties of the single-particle basis and the reference state result from a choice such that various combinations of $G_\text{bas}$ and $G_\text{ref}$ can be employed.

\subsection{Symmetries of the single-particle basis}

Given $H$ and its symmetry group $\Gham$, there is infinitely many different single-particle bases spanning the one-body Hilbert space $\mathcal{H}_1$ that can be used to represent the operator in second-quantized form. The single-particle basis functions are typically obtained as eigenstates of an auxiliary one-body Hamiltonian $H_\text{bas}$ whose symmetries are characterized by
\begin{align}
[H_\text{bas} , U(g)] = 0 \quad (\forall g \in \Gbas) .
\end{align}
When choosing $\Gbas=SU(2)$, one-body basis states are eigenstates of the squared total angular-momentum operator\footnote{Vectors are represented in bold face.}
\begin{align}
\vec{J}^2 \equiv J_x^2 + J_y^2 + J_z^2\, ,
\end{align}
where $J_x,J_y$ and $J_z$ denote the Cartesian components of the total angular-momentum vector. In most \emph{ab initio} nuclear structure applications such a one-body basis is indeed employed, e.g., the eigenbasis of the three-dimensional spherical harmonic oscillator (sHO) Hamiltonian
\begin{align}
H_\text{sHO} \equiv \frac{\vec{p}^2}{2m} + \frac{1}{2} m \omega^2 \vec{r}^2\, ,
\label{eq:HOham}
\end{align}
where $m$ denotes the average nucleon mass and $\omega$ the HO frequency. It can be shown that
\begin{subequations}
\begin{align}
 [H_{\text{sHO}}, \vec{J}^2] =0 \, , \\
 [H_{\text{sHO}}, J_z]  =0 \, ,
\end{align}
\end{subequations}
such that the one-body eigenstates of $H_{\text{sHO}}$ are proportional to spherical Harmonics. In other frameworks, e.g. nuclear energy density functional calculations, the single-particle basis is possibly taken as eigenfunctions of the axially deformed HO Hamiltonian that breaks rotational invariance and, thus, displays a smaller symmetry group $\Gbas$ than $H_{\text{sHO}}$.

\subsection{Symmetries of the reference state}

The rationale of expansion methods relies on the definition of a conveniently chosen $A$-body reference state $\ket{\Phi }$ that serves as starting point for the correlation expansion. Acting on the vacuum, the wave operator $W$ yields the exact, e.g., ground state
\begin{align}
\ket{\Psi_0 } = W \ket{ \Phi } \, .
\end{align}
The wave operator is expanded and truncated according to a given many-body scheme, e.g., in MBPT, SCGF or CC theory. The resulting equations are symmetry-unrestricted and therefore make no use of symmetry properties of many-body operators.

In practice, the reference state is typically obtained as the ground state of an 'unperturbed' Hamiltonian $H_\text{ref}$ capturing the average behavior of the system's dynamics and characterized by a symmetry group $\Gref$ \begin{align}
[ H_\text{ref} , U(g) ] = 0  \quad (\forall g \in \Gref) \, ,
\end{align}
such that $\ket{\Phi}$ typically belongs to the trivial IRREP of $\Gref$. In the following, the reference state $\ket{\Phi_{\Gref} }$, thus, carries a subscript specifying the symmetry group of the Hamiltonian it is the ground state of.

In the simplest case, the vacuum is chosen to be a Slater determinant $\ket{\Phi_{\Gham} }$ obtained from a symmetry-restricted Hartree-Fock mean-field calculation, i.e.
\begin{align}
\Gref = \Gham \, .
\end{align}
In nuclear systems, $\ket{\Phi_{\Gham} }$ typically belongs to the trivial IRREP of $SU(2)$ and $U(1)$, i.e., it carries good angular momentum $J=0$ and a fixed number of particles. Dynamic correlations are introduced via the action of the wave operator that generates summations over elementary particle-hole excitations.

In open-shell systems, the above reference state is improper due to the partial filling of the last occupied shell. This leads to a degeneracy with respect to particle-hole excitations, thus, signalling the existence of a Goldstone mode and the ill-definition of the previously performed expansion of $W$. This problem can be circumvented by lowering the symmetry group of $H_\text{ref}$, i.e. by taking a well-chosen subgroup $\Gref \subset \Gham $. This typically leads to breaking $U(1)$ symmetry in singly open-shell nuclei and/or $SU(2)$ in doubly open-shell nuclei. The lower symmetry of $H_\text{ref}$ induces a lower reference energy due to the enlarged variational space
\begin{align}
E[\ket{\Phi_{\Gref} } ]  \leq E[ \ket{ \Phi_{\Gham}}] \, .
\end{align}
More importantly, this lowering is accompanied by a lifting of the degeneracy of $\ket{\Phi_{\Gham} }$ with respect to elementary excitations such that $W$ can be expanded safely. In this case, however, the wave operator must not only capture dynamical correlations but also restore the symmetry  ${\Gham}$ associated with the exact eigenstates of $H$. Because of the necessary truncation, a standard expansion of $W$ is not capable of restoring the symmetry such that the symmetry contamination needs to be retrieved by the explicit inclusion of a symmetry projector in the definition of $W$~\cite{Duguet:2014jja,Duguet:2015yle,Qiu:2019edx}.

\subsection{Reduction schemes and groups}

While the symmetry group of the Hamiltonian is fixed from the outset, the choice of the single-particle basis and reference states leaves tremendous freedom to adapt $\Gbas$ and $\Gref$ in order to deal with a specific situation.

A case of particular interest arises when the symmetry groups of the Hamiltonian, the single-particle basis and the reference state coincide, i.e.,
\begin{align}
 \Gsym  \equiv\Gham = \Gbas = \Gref \, .
\end{align}
In this setting, the common algebraic structure can be exploited to simplify the many-body formalism by expressing all working equations in terms of $\Gsym$-reduced tensors, thus, potentially providing a tremendous gain in the required runtime and memory resources. In the present paper, this situation is exploited relative to the $SU(2)$ group (independently of the treatment of other symmetries such as $U(1)$).

\subsection{Tensors and tensor networks}

Due to the large variety of expansion schemes built to retrieve the solution of the many-body problem, it is desirable to introduce a unifying language for the various frameworks.
This common ground is provided by the language of tensors and tensor networks.

A mode-$k$ \emph{symmetry-unrestricted tensor} (SU-T)
\begin{align}
T_{i_1 ... i_k}
\end{align}
is a multi-variate data array carrying $k$ indices with (possibly different) index ranges $I_1,...,I_k$. Tensors constitute the basic building blocks of many-body expansion methods. Given a set of SU-T's $A,B,C,...$ a \emph{contraction} is defined as a summation over a common index, e.g.,
\begin{align}
\sum_{k} A_{...k...} B_{...k...} C_{...k...}\, , \nonumber
\end{align}
where the ellipses indicate indices that are not summed over\footnote{Indices may appear more than twice, a feature uncommon for traditional contractions as in the theory of general relativity.}.

A \emph{symmetry-unrestricted tensor network} (SU-TN) denotes a set of SU-T's combined according to a given \emph{contraction scheme} specifying the way the tensors are contracted with each other. Furthermore, a SU-TN is said to be \emph{closed} if all tensor indices are summed over and is said to be \emph{open} otherwise.

In many-body applications tensors typically appear in two broad classes
\begin{aenumerate}
\item \emph{input tensors} that are known prior to addressing the actual solution of the Schr\"odinger equation in a given many-body framework,
\item \emph{output tensors} that are specific to a given many-body approach and are typically the objects being solved for.
\end{aenumerate}
Examples for input tensors are matrix elements of many-body operators like the Hamiltonian whereas examples of output tensors are CC amplitudes or dressed propagators in SCGF theory. In most non-perturbative many-body frameworks, like CC, IMSRG or SCGF, open TN's specify the working equations required to determine the unknown output tensors while the calculation of observables, e.g. the energy, relates to the evaluation of closed TNs.

\subsection{Symmetry-reduced tensor networks}

The goal of this work is to transform an initial SU-TN into a \emph{symmetry-reduced tensor network} (SR-TN) encapsulating the symmetry reduction according to the associated symmetry group. To do so, the SU-T's must be replaced by their symmetry-reduced counterparts. Given an initial SU-T, the corresponding \emph{symmetry-reduced tensor} (SR-T) is obtained from a transformation $f_{\Gsym}$
\begin{align}
T_{k_1...k_n} \xrightarrow{ \phantom{n} f_{\Gsym} \phantom{n} }   \tilde T^{\lambda}_{\tilde k_1 ... \tilde k_n} \, ,
\end{align}
mediating the symmetry reduction related to the group $\Gsym$.
Here, the symbol $\lambda$ denotes the relevant IRREP labels of the symmetry-reduced tensor.
In the following, quantities with a tilde indicate symmetry-reduced objects. Note that the content of the indices themselves change, such that the set of quantum numbers labelling a SU-T and its SR-T counterpart are different. Thus, the SR-TN denotes the end product obtained via the replacement of the SU-T's by their SR-T counterparts and via the adjustment of the contraction pattern
\begin{equation}
  \sum_{k} A_{...k...} B_{...k...} C_{...k...}  \xrightarrow{ \mspace{8mu} f_{\Gsym} \mspace{8mu} }  \sum_{\lambda \tilde k } \tilde A^{\lambda}_{...\tilde k...} \tilde B^{\lambda}_{...\tilde k...} \tilde C^{\lambda}_{...\tilde k...}\, . \nonumber
\end{equation}

\section{Angular-momentum algebra}
\label{sec:ama}

\subsection{Rationale}
\label{rationale}

While the discussion on symmetry-reduction and SR-TN's has been generic so far, the present paper focus on the $SU(2)$ group. The goal is, thus, to obtain \emph{angular-momentum-re\-duced tensor networks} (AMR-TN's) from SU-TN ones. The procedure requires to
\begin{aenumerate}
\item replace all the SU-T's by their AMR-T counterparts according to the transformation $f_{SU(2)}$,
\item constrain the contraction pattern to only be left with summations over the reduced set of quantum numbers.
\end{aenumerate}
In practice, step (1) involves a set of \emph{substitution rules} for every many-body tensor at play that specify how the symmetry reduction is performed. The resulting SR-TN---and its computational complexity---may strongly depend on the choice made to perform this initial step\footnote{This step is not uniquely defined as several choices for the same group can be envisioned.}. From this point of view at least a minimum level of human input (and experience) is necessary to come up with the most convenient choice. This does not pose a severe limitation in any of the examples discussed below.

\subsection{Other symmetries}

While presently focusing on rotational symmetry, other symmetries can be exploited in the same way. A key example relates to intrinsic spin in quantum chemistry that is analogous to the total angular-momentum when using a $ls$-coupling scheme. The spin projection being only two-fold degenerate, i.e.\ $m_s = \pm \frac{1}{2}$, spin-restricted many-body theories benefit less from the symmetry reduction than in the $j$-coupling scheme. Still, pre-processing the sums over spin projections is an important tool to reduce the computational cost and advance state-of-the-art expansion methods in strongly correlated electronic systems. Finite symmetry groups, e.g., the dihedral groups $D_n$, may also arise in quantum molecules whereas cubic groups play an important role in the computation of homogeneous matter, e.g., the infinite electron gas or infinite nuclear matter, since periodic boundary conditions are employed to facilitate the calculation. In solid-state physics, symmetry properties of the many-body systems, e.g., helical symmetries in nano tubes, can also be exploited to reduce computational complexity.

All the aforementioned examples correspond to a reduction of exact symmetries of a many-body system.
In recent years, exploiting emergent approximate symmetries has also been shown to be highly beneficial, in particular in the context of nuclear CI-based approaches. In this case, the symmetry group of the configuration basis $\Gbas$ is \emph{larger} than the actual symmetry group of the Hamiltonian,
\begin{align}
\Gham \subset \Gbas \, ,
\end{align}
thus, exploiting algebraic properties that are not strictly realized in nature. A prime example is the symplectic symmetry group $Sp(3,\mathbb{R})$ that is not an exact symmetry of the nuclear Hamiltonian but of the kinetic energy operator.
In the symmetry-adapted no-core shell model (SA-NCSM) an $A$-body configuration basis is constructed from the Casimir operators of the approximate symmetry group $SU(3)\subset Sp(3,\mathbb{R})$. The use of symplectic algebra was shown to provide an efficient selection of many-body basis states, thus, yielding computational savings in the diagonalization of the many-body Hamiltonian at the price of a more involved hand{\-}ling of many-body operators~\cite{Dytrych2008}.

\subsection{$SU(2)$ group}
\label{su2}

In order to move closer to a concrete implementation of the above procedure, let us introduce details about the nonabelian compact  $SU(2)\equiv \{R(\Omega), \Omega \in  D_{SU(2)}\}$ Lie group associated with the rotation of a $A$-body fermion system characterized by an integer or a half-integer angular momentum. The group is parametrized by three Euler angles $\Omega \equiv (\alpha,\beta,\gamma)$ whose domain of definition is
\begin{equation}
D_{SU(2)} \equiv D_{\alpha} \times D_{\beta} \times D_{\gamma}  = [0,4\pi] \times [0,\pi] \times [0,2\pi] \, .
\end{equation}
As $SU(2)$ is considered to be a symmetry group of $H$, the commutation relations
\begin{align}
\left[H,R(\Omega)\right] = \left[T,R(\Omega)\right] = \left[V,R(\Omega)\right] =0 \, , \label{commutation}
\end{align}
hold for $\Omega \in  D_{SU(2)}$.

Subsequently, the unitary representation of $SU(2)$ on Fock space is utilized
\begin{equation}
R(\Omega)  =e^{-\frac{i}{\hbar}\alpha J_{z}}e^{-\frac{i}{\hbar}\beta J_{y}}e^{-\frac{i}{\hbar}\gamma J_{z}} \, .
\end{equation}
The components of the total angular-momentum vector make up the Lie algebra
\begin{equation}
[J_{i},J_{j}]=\epsilon_{ijk} i\hbar \, J_{k} \, , \label{Lieidentity}
\end{equation}
where $\epsilon_{ijk}$ denotes the Levi-Civita tensor. The \emph{Casimir operator} of the group built from the infinitesimal generators through a non-degenerate invariant bilinear form is the total angular momentum
\begin{equation}
\vec{J}^{2}  \equiv \sum_{i=x,y,z} J^2_{i} \, .
\end{equation}

Matrix elements of the irreducible representations (IRREPs) of $SU(2)$ are given by the so-called Wigner $D$ functions~\cite{VaMo88}
\begin{equation}
\braket{ \xi J M |  R(\Omega)  | \xi' J' M'  } \equiv \delta_{\xi\xi'} \delta_{JJ'} D_{MM'}^{J}(\Omega) \, ,
\end{equation}
where $\ket{ \xi J M }$ is an eigenstate of $\vec{J}^2$ and $J_{z}$
\begin{subequations}
\label{eigenequationJ}
\begin{align}
\vec{J}^2 \ket{ \xi J M  } &=  J(J+1)\hbar^2 \ket{ \xi J M } \, , \\
J_{z} \ket{ \xi J M } &=  M\hbar \ket{ \xi J M } \, ,
\end{align}
\end{subequations}
with $2J \in \mathbb{N}$, $2M\in \mathbb{Z}$, $J-M\in \mathbb{N}$ and $-J\leq M \leq +J$. The index $\xi$ collects all quantum numbers but $J$ and $M$.  The $(2J\!+\!1)$-dimensional IRREPs are labelled by $J$ and are spanned by the set of states $\{\ket{ \xi J M }\}$ for fixed $J$ and $\xi$.

An irreducible tensor operator ${\mathbf{T}^{J}}$ of rank $J$ is made of $2J+1$ operators  $T^{J}_{K}$ transforming under rotation as
\begin{subequations}
\label{eq:ten:def}
\begin{align}
R(\Omega) \, T^{J}_{K} \, R(\Omega)^{-1}  &= \sum_{M}  T^{J}_{M} \, D^{J}_{MK}(\Omega) \,\,\, , 
\end{align}
\end{subequations}
or, equivalently, fulfilling
\subalign{
[J_z, T^{J}_{K}] &= \hbar K \, T^{J}_{K} \, ,\\
[J_\pm, T^{J}_{K} ] &= \hbar \sqrt{(J \pm K + 1)(J\mp K)} \, T^{J}_{K\pm1} \, ,
}
where $J_\pm = J_x \pm i J_y$ denotes the usual raising (lowering) operators. The nuclear Hamiltonian is an example of a spherical tensor operator of rank zero. Such operators are denoted as \emph{scalar}.

A powerful tool to treat spherical tensor operators is the celebrated \emph{Wigner-Eckart theorem} (WET)
\begin{align}
\MoveEqLeft\braket{ \xi_1 j_1 m_1 | T^{J}_{M} | \xi_2 j_2 m_2 } \notag\\
&= (-1)^{2J} \frac{1}{\jhat{\jmath}_1} \clebsch{j_2}{m_2}{J}{M}{j_1}{m_1}
( \xi_1 j_1  | {\mathbf{T}^{J}} | \xi_2 j_2  )\, ,
\label{eq:WEthm}
\end{align}
where $\jhat{\jmath}\equiv \sqrt{2j+1}$. The theorem states that matrix elements of a given component of a spherical tensor in the basis spanning the IRREPs can be written as a product of a \emph{geometric part} independent of the spherical tensor at play and of a \emph{reduced matrix element} independent on the particular component of the spherical tensor and the members of the IRREPs under consideration~\cite{Su07}.

In the special case of a scalar operator one has
\begin{align}
\braket{ \xi_1 j_1 m_1 | T^{0}_{0} | \xi_2 j_2 m_2 } = \frac{1}{\jhat{\jmath}_1} ( \xi_1 j_1  | {\mathbf{T}^{0}} | \xi_2 j_2  )\, ,
\label{eq:WEthm scalar}
\end{align}
such that both the initial and reduced matrix elements are independent of any projection quantum number. In this particular case, the notion of reduced matrix element is thus irrelevant.

\subsection{Fermionic algebra}

One of the building blocks of quantum many-body theory are second-quantized operators
\begin{align}
O^{ij} = \frac{1}{i!j!} \sum_{k_1 ... k_{i+j}} \hspace{-2pt}\bar o_{k_1 ... k_{i+j}} c^\dagger_{k_1} \cdots c^\dagger_{k_i} c_{k_{i+j}} \cdots c_{k_{i+1}} \, ,
\end{align}
where $c^\dagger$ ($c$) denote single-particle creation (annihilation) operators associated with a basis $\mathcal{B}_1$ of the one-body Hilbert space
\begin{align}
\mathcal{H}_1\equiv \mathcal{H}^{r}_1 \otimes \mathcal{H}^{s}_1 \otimes \mathcal{H}^{t}_1
\end{align}
that is the tensor product of a spatial part, a spin part and an isospin part. Anti-symmetrized matrix elements $\bar o_{k_1 ... k_{i+j}}$  carrying $(i+j)$ one-body indices constitute a mode-$(i+j)$ SU-T. Creation and annihilation operators are assumed to fulfil the canonical anti-commutation rules
\begin{subequations}
\begin{align}
\{c_{k_1},c_{k_2}\} &= 0 \, , \\
\{c^\dagger_{k_1},c^\dagger_{k_2}\} &= 0 \, ,\\
\{c_{k_1},c^\dagger_{k_2}\} &= \delta_{k_1 k_2} \, ,
\end{align}
\end{subequations}
defining the Fermionic algebra\footnote{When breaking $U(1)$ symmetry, one employs the quasi-particle algebra associated with the Bogoliubov transformation~\cite{RiSc80}
\begin{align}
\beta_k &\equiv \sum_p U^*_{pk} c_p + V^*_{pk} c^\dagger_p, &
\beta_k^\dagger &\equiv \sum_p U_{pk} c^\dagger_p + V_{pk} c_p \, ,
\end{align}
such that operators are expressed in this basis and that the indices of the associated matrix elements relate to quasi-particles. This feature does not change fundamentally what follows regarding the handling of $SU(2)$ symmetry.}. Processing many-body matrix elements of strings of such operators via various forms of Wick's theorem is at the core of quantum many-body methods and gives rise to the multitude of TN's at play in expansion methods.

\subsection{$SU(2)$ symmetry and basis states}

In the following, $\mathcal{B}_1$ is taken to be the eigenbasis of a $SU(2)$-invariant Hamiltonian  $H_\text{bas}$ such that basis states are conveniently labeled as
\begin{align}
\ket{ k }  = \ket{ n_k l_k j_k m_{j_k} t_k }\, , \label{defB1}
\end{align}
where $n_k$ denotes the radial quantum number, $l_k$ the orbital quantum number, $j_k$ the total angular-momentum quantum number, $m_{j_k}$ its projection and $t_k$ the isospin projection. This constitutes a so-called $j$-coupled basis, i.e. it is not a direct product of bases of $\mathcal{H}^{r}_1$ and $\mathcal{H}^{s}_1$ but a coupled basis whose members are eigenstates of the total angular momentum $\vec{j}^2$. While the eigenstates of the aforementioned sHO Hamiltonian provide a  example of practical interest, eigenbases of other one-body Hamiltonians characterized by rotational symmetry are equally valid.

Later on, the AMR-T's employed throughout the symmetry reduction will carry reduced labels $\tilde k$ characterized by
\begin{align}
\tilde k  \equiv  (n_k , l_k, j_k, t_k) \, ,
\end{align}
where the angular-momentum projection, i.e. the magnetic quantum number $m_k$, is explicitly excluded compared to the definition of $k$ through the given of the basis in Eq.~\eqref{defB1}.

The tensor product of two one-body states defines a basis state of the two-body Hilbert space ${\cal H}_2$
\begin{align}
\ket{ k_1 k_2 } \equiv \ket{k_1 } \otimes \ket{k_2 } \, .
\end{align}
Contrarily, in the \emph{coupled representation} the two total angular momenta $j_{k_1}$ and $j_{k_2}$ are coupled to a total two-body angular momentum $J$ with projection\footnote{While coupled two-body states do indeed depend on $M$, the label is omitted for brevity given that the $M$-dependence of the reduced tensors is completely specified by the WET.} $M$,
\begin{align}
\ket{ \tilde k_1 \tilde k_2 (J) } \equiv \sum_{m_{k_1} m_{k_2}} \clebsch{j_{k_1}}{m_{k_1}}{j_{k_2}}{m_{k_2}}{J}{M} \, \ket{k_1 k_2 } \, ,
\label{eq:twobJ}
\end{align}
where the vector space inner product
\begin{align}
\clebsch{j_{k_1}}{m_{k_1}}{j_{k_2}}{m_{k_2}}{J}{M} \equiv \braket{ k_1 k_2 | \tilde k_1 \tilde k_2 (J) }
\end{align}
denotes the Clebsch-Gordan (CG) coefficient mediating the transformation from the uncoupled to the coupled basis. The left-hand side of Eq.~\eqref{eq:twobJ} defines two-body eigenstates of $\op{J^2}$, the Casimir operator of the group. The inverse transformation of Eq.~\eqref{eq:twobJ} is given by
\begin{align}
\ket{  k_1 k_2 } = \sum_{J=|j_{k_1} -j_{k_2} |}^{j_{k_1} +j_{k_2}} \sum_{M=-J}^J \clebsch{j_{k_1}}{m_{k_1}}{j_{k_2}}{m_{k_2}}{J}{M} \, \ket{ \tilde k_1 \tilde k_2 (J) } \, .
\end{align}

Along the same lines, the uncoupled three-body basis states of  ${\cal H}_3$, i.e., the tensor product of three single-particle states
\begin{align}
\ket{ k_1 k_2 k_3 } \equiv \ket{k_1 } \otimes \ket{k_2 } \otimes \ket{k_3 }
\end{align}
is defined.
Performing the angular-momentum coupling requires fixing the coupling order which is subsequently chosen to be
\begin{align}
  &\ket{ [\tilde k_1 \tilde k_2 (J_{12})] \tilde k_3 (J) } \notag \\
  &= \sum_{ \substack{m_{k_1} m_{k_2} \\ m_{k_3} M_{12}}}
  \clebsch{j_{k_1}}{m_{k_1}}{j_{k_2}}{m_{k_2}}{J_{12}}{M_{12}}
  \clebsch{J_{12}}{M_{12}}{j_{k_3}}{m_{k_3}}{J}{M}  \ket{ k_1 k_2 k_3 } \, ,
\label{eq:threebJ}
\end{align}
i.e., the first two single-particle states are coupled to an intermediate two-body angular-momentum quantum number $J_{12}$, which is further coupled to the third state to yield the overall (half-integer) three-body angular-momentum $J$. In analogy to the two-particle case, Eq.~\eqref{eq:threebJ} defines three-body eigenstates of $\vec{J}^2$.

The choice of the coupling order for three-body states employed in Eq.~\eqref{eq:threebJ} is arbitrary such that an alternative coupling scheme is given by
\begin{align}
&\ket{ [\tilde k_1 [\tilde k_2  \tilde k_3(J_{23})] (J) }  \notag \\
  & = \sum_{ \substack{m_{k_1} m_{k_2} \\ m_{k_3} M_{12}}}
\clebsch{j_{k_2}}{m_{k_2}}{j_{k_3}}{m_{k_3}}{J_{23}}{M_{23}}
\clebsch{J_{23}}{M_{23}}{j_{k_1}}{m_{k_1}}{J}{M}  \ket{ k_1 k_2 k_3 } \, , 
\label{eq:3b2}
\end{align}
where the second and third single-particle states are coupled to an intermediate angular momentum $J_{23}$ that is subsequently coupled with $j_{k_1}$ to an overall $J$. Both coupling schemes enable for the construction of a basis of  ${\cal H}_3$ that is an eigenbasis of $\vec{J}^2$. The transformation between the two representations is given by
\begin{align}
\MoveEqLeft\ket{ [\tilde k_1 \tilde k_2 (J_{12})] \tilde k_3 (J) } \notag \\
&= (-1)^{j_1+j_2+j_3+J}\sum_{J_{23}} \hat{J}_{12}\hat{J}_{23} \sixj{j_1}{j_2}{J_{12}}{j_3}{J}{J_{23}}
\notag\\&\phantom{{}={}}\times
\ket{ \tilde k_1 [\tilde k_2 \tilde k_3 (J_{23})] (J) } \, ,
\label{eq:lindepm}
\end{align}
where the Wigner $6j$-symbol was introduced.

Recursively, $N$-body states can be introduced for $N\geq 3$, e.g. for the uncoupled representation
\begin{align}
\ket{k_1 \cdots k_N } \equiv \bigotimes_{i=1}^N \ket{k_i } \, .
\end{align}
Since in current \emph{ab initio} implementations four- and higher-body operators play no dominant role yet, this extension is not discussed here.

\subsection{Many-body matrix elements}

With the operator $O$ being the $\mu$ component of a spherical tensor of rank $\lambda$, its uncoupled matrix elements are defined by
\begin{align}
  \bar o_{k_1 ...k _i k_{i+1} ... k_{i+j}} \equiv \braket{ k_1 ... k_i | O^{\lambda}_{\mu} | k_{i+1} ... k_{i+j} } \, ,
\end{align}
where it is not assumed that the numbers of indices labelling the bra and the ket states coincide. By means of the transformations between uncoupled and coupled representation of the bra and ket states, coupled expressions for matrix elements can be derived.
Focusing on a two-body operator characterized by uncoupled matrix elements $\bar o_{k_1 k_2 k_3 k_4}$, their angular-momentum-coupled counterparts are\footnote{The transformation can in principle be accompanied by an additional phase factor.}
\begin{align}
  \tilde O^{JMJ^\pr M^\pr}_{\tilde k_1 \tilde k_2 \tilde k_3 \tilde k_4} &= \sum_{ \substack{m_{k_1} m_{k_2} \\ m_{k_3} m_{k_4} }}  \bar o_{k_1 k_2 k_3 k_4}   \clebsch{j_{k_1}}{m_{k_1}}{j_{k_2}}{m_{k_2}}{J}{M} \clebsch{j_{k_3}}{m_{k_3}}{j_{k_4}}{m_{k_4}}{J^\pr}{M^\pr} \, ,
 \label{eq:twobodyJ}
\end{align}
Analogously, coupled three-body matrix elements are obtained as
\begin{align}
\MoveEqLeft\tilde O^{J_{12}JM J_{45} J^\pr M^\pr}_{\tilde k_1 \tilde k_2 \tilde k_3 \tilde k_4 \tilde k_5 \tilde k_6} \notag \\
&=\smashoperator[r]{\sum_{ \substack{m_{k_1} m_{k_2} m_{k_3} M_{12} \\ m_{k_4} m_{k_5} m_{k_6} M_{45}}}}
\quad \quad \clebsch{j_{k_1}}{m_{k_1}}{j_{k_2}}{m_{k_2}}{J_{12}}{M_{12}}
\clebsch{J_{12}}{M_{12}}{j_{k_3}}{m_{k_3}}{J}{M} \notag \\ &\phantom{{}={}}\times
\clebsch{j_{k_4}}{m_{k_4}}{j_{k_5}}{m_{k_5}}{J_{45}}{M_{45}}
\clebsch{J_{45}}{M_{45}}{j_{k_6}}{m_{k_6}}{J^\pr}{M^\pr} \notag \\ &\phantom{{}={}}\times
\bar o_{k_1k_2 k_3 k_4 k_5 k_6}\, .
\label{eq:threebodyJ}
\end{align}
Neither Eq.~\eqref{eq:twobodyJ} nor Eq.~\eqref{eq:threebodyJ} assume the underlying operator to be scalar.
If it is indeed the case, the selection rules $J=J^\pr$ and $M=M^\pr$ hold and the coupled matrix element is additionally independent of $M$. For the three-body operator, however, the intermediate couplings do not necessarily coincide, i.e., $J_{12} \neq J_{45}$ in general.

\section{Diagrammatic method}
\label{sec:amcdiag}

Even though all manipulations necessary to simplify angular-momentum expressions can be performed solely in terms of the expressions introduced in Sec.~\ref{sec:ama} it is at the heart of this work to introduce a more convenient representation of the involved algebraic steps that, additionally, allows for computer-aided derivations. As Feynman or Goldstone diagrams are used to efficiently capture the results of cumbersome applications of Wick's theorem, diagrams can be introduced to restate complicated identities associated with angular momentum algebra~\cite{VaMo88}. A modern account of the underlying group-theoretic properties is provided in Ref.~\cite{Wormer2006}. For completeness, let us mention that similar frameworks can be introduced to tackle other (more involved) symmetry groups\footnote{Groups (e.g. $SU(n)$, $n>2$) for which a given IRREP may appear several times in the process of reducing the product of two IRREPs require the introduction of a \emph{multiplicity label} for each IRREP.}. The interested reader is referred to Ref.~\cite{Stedman1990} for an extensive discussion.

\subsection{Preliminaries}

As seen in Sec.~\ref{sec:ama}, CG coefficients constitute the basic building blocks of angular-momentum theory. However, CG coefficients are somewhat inconvenient due to their asymmetry with respect to the involved angular-momenta.  A more symmetric representation can be obtained in terms of \emph{Wigner $3jm$-symbols}\footnote{Although the $3jm$-symbols are usually referred to as $3j$-symbols in the literature, this terminology is used here in order to distinguish them from the $3j$-, $6j$- and $9j$-symbols appearing in the remainder of this document.}
\begin{align}
\threej{j_1}{m_1}{j_2}{m_2}{j_3}{m_3} \equiv \frac{1}{\jhat{j}_1} (-1)^{j_2 -j_3 -m_1} \clebsch{j_1}{-m_1}{j_2}{m_2}{j_3}{m_3} \, .  \label{identity1}
\end{align}
Wigner $3jm$-symbols are invariant under cyclic column permutations,
\begin{align}
\threej{j_1}{m_1}{j_2}{m_2}{j_3}{m_3} =  \threej{j_3}{m_3}{j_1}{m_1}{j_2}{m_2} = \threej{j_2}{m_2}{j_3}{m_3}{j_1}{m_1} \, ,  \label{identity2}
\end{align}
whereas anti-cyclic permutations induce a phase factor
\begin{align}
\threej{j_1}{m_1}{j_2}{m_2}{j_3}{m_3} =  (-1)^{j_1 + j_2 + j_3 } \threej{j_2}{m_2}{j_1}{m_1}{j_3}{m_3} \, . \label{identity3}
\end{align}
Wigner $3jm$-symbols with opposite magnetic quantum numbers are related via the identity
\begin{align}
\threej{j_1}{m_1}{j_2}{m_2}{j_3}{m_3} =  (-1)^{j_1 +j_2+j_3} \threej{j_1}{-m_1}{j_2}{-m_2}{j_3}{-m_3} \, . \label{identity4}
\end{align}
Furthermore, $3jm$-symbols with one vanishing $(j,m)$ pair simplify according to
\begin{align}
\threej{j_1}{m_1}{j_2}{-m_2}{0}{0}  = (-1)^{j_1 -m_1} \frac{1}{\jhat{j}_1} \delta_{j_1 j_2} \delta_{m_1 m_2} \, .
\label{eq:zero line}
\end{align}

\subsection{Vertices}

Wigner $3j$-symbols provide the building blocks of the diagrammatic formalism. They are represented by vertices in the so-called Yutsis graphs.%
\footnote{Named after Lithuanian Adolfas Jucys, these are also known as Jucys graphs.
Since many of his works were initially published in Russian, the transliteration of the name from Russian, Yutsis graph, is the better-known one.}
More specifically, a vertex carrying three outgoing lines, each labelled by a tuple $(j_k,m_k)$, represents the $3j$-symbol
\diagram{3j_def}
The vertex sign denotes a convention specifying the column order that must be used to write the corresponding $3jm$-symbol, i.e., a plus (minus) sign stipulates that the lines and the associated angular-momentum labels must be read counterclockwise (clockwise).

Furthermore, the vertex with one ingoing line represents
\diagram{3j_invline3j}

Starting from the two above definitions, the vertices with two and three ingoing lines are obtained by applying the operation consisting of inverting the directions of all three lines at once. Starting for example from the vertex with three outgoing lines, one obtains the vertex with three ingoing lines
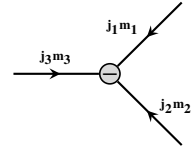
\diagram{3j_rev}
whose expression is given by
\begin{align}
(-1)^{j_1-m_1 +j_2-m_2+j_3-m_3} \threej{j_1}{-m_1}{j_2}{-m_2}{j_3}{-m_3} \, ,
\end{align}
as a testimony of Eq.~\eqref{identity4} and where the magnetic quantum numbers have been added to the phase at no cost given that $m_1+m_2+m_3=0$ holds. Through this operation, the sign is not altered. Additionally, performing the operation twice does give back the original vertex thanks to the identity
\begin{align}
(-1)^{2(j_1-m_1 +j_2-m_2+j_3-m_3)} =1 \, .
\end{align}

Changing the sign carried by the vertex can be performed at the price of the phase factor
\begin{align}
\Phi_\text{ns} = (-1)^{j_1 + j_2 + j_3}\, ,
\end{align}
where the lower index 'ns' stipulates the \emph{node sign reversal}.
Indeed, moving from a clockwise to a counterclockwise (or vice versa) reading of the vertex corresponds to performing one column inversion in the  $3j$-symbol whose effect is characterized by Eq.~\eqref{identity3}. Notice that changing the vertex sign is equivalent to moving one line across another one.

\subsection{Yutsis graphs}

The network of  $3jm$-symbols generated via step (1) of the angular-momentum reduction of a SU-TN (see Sec.~\ref{rationale}) is represented by a Yutsis graph. Those graphs are, thus, obtained by contracting a set of vertices through their edges in a way that consistently represent the network of  $3j$-symbols.

Contracting the edges of two vertices is possible if both lines carry the same angular momentum quantum numbers $(j,m)$ and go in the same direction, i.e., one must be going out of the first vertex while the other one must be going into the second vertex
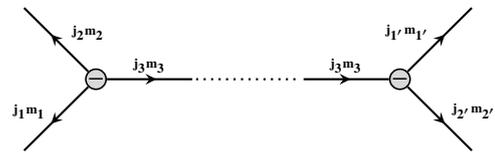
\diagram{contraction_def}
The contraction itself corresponds to summing over the common magnetic quantum number such that the internal line does not carry it anymore
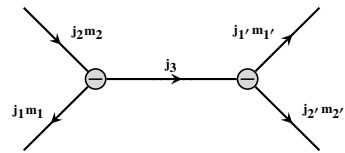
\diagram{contraction_network}
Reading the vertices according to the definitions given previously, the algebraic expression resulting from the contraction reads as
\begin{align}
\sum_{m_3} (-1)^{j_3-m_3} \threej{j_1}{m_1}{j_2}{m_2}{j_3}{m_3}  \threej{j_{1^\prime}}{m_{1^\prime}}{j_{2^\prime}}{m_{2^\prime}}{j_3}{-m_3} \, .
\end{align}
Given a Yutsis graph, the direction of an internal line carrying angular momentum $j$ can be reversed at the price of accounting for the phase factor
\begin{align}
\Phi_\text{rev} = (-1)^{2j}\, .
\end{align}
An example of practical interest relates to fully contracting the two vertices
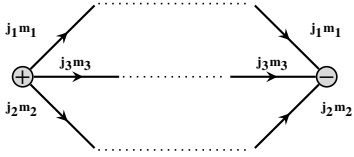
\diagram{tridelta_contraction}
to generate the closed Yutsis graph
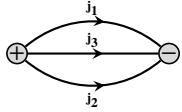
\diagram{tricond}
actually corresponding to the so-called \emph{Wigner $3j$-symbol} $\trij{j_1}{j_2}{j_3}$, also called \emph{triangular delta}\footnote{Some texts call the triangular delta a $3j$-symbol, as it technically is the first of the $3nj$-symbols, while naming the coupling coefficients $3jm$-symbols. This nomenclature in is not adopted in order to avoid confusion.} or \emph{triangular inequality}. The corresponding algebraic expression is given by
\begin{align}
\trij{j_1}{j_2}{j_3} & = \sum_{m_1 m_2 m_3}  (-1)^{j_1-m_1+ j_2-m_2+j_3-m_3}  \notag \\ &\times \threej{j_1}{m_1}{j_2}{m_2}{j_3}{m_3} \threej{j_{1}}{-m_{1}}{j_2}{-m_{2}}{j_3}{-m_3} \nonumber \\
&= \begin{cases} 1, \quad \text{if} \, |j_1 - j_2| \leq j_3 \leq j_1 + j_2 \\ 0 , \quad \text{otherwise} \end{cases} \, ,
\end{align}
which vanishes unless the inequalities are satisfied.

\subsection{Unfactorizable graphs}
\label{unfactorizable}

Wigner $3nj$-symbols provide relevant examples of Yutsis graphs that cannot be simplified via factorization rules.  The first example is the \emph{Wigner $6j$-symbol} that is graphically represented as a tetrahedral structure
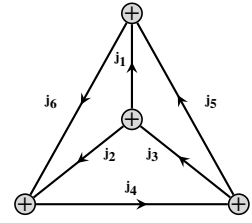
\diagram{6j_triangle}
Translating the central vertex to the upper right corner and accounting for the change in the ordering of the lines attached to the upper-left and the lower-right vertices, the diagram can be equally represented as a square with two diagonal lines
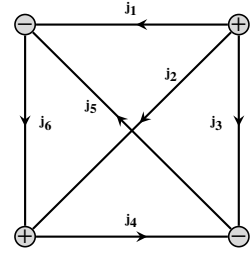
\diagram{6j_square}
Independently of which of the two diagrams is used, the corresponding algebraic expression is
\begin{align}
\sixj{j_1}{j_2}{j_3}{j_4}{j_5}{j_6} &\equiv \sum_{m_1 ... m_6}
(-1)^{\sum_{k=1}^{6} (j_k -m_k)} \notag \\ &\phantom{{}={}} \times
\threej{j_1}{m_1}{j_2}{m_2}{j_3}{m_3} \threej{j_1}{-m_1}{j_5}{-m_5}{j_6}{m_6} \notag \\
&\phantom{{}={}} \times \threej{j_4}{m_4}{j_2}{-m_2}{j_6}{-m_6} \threej{j_4}{-m_4}{j_5}{m_5}{j_3}{-m_3}
\end{align}

The case $n=3$ yields the Wigner $9j$-symbol whose algebraic expression
\begin{align}
\ninej{j_1}{j_2}{j_3}{j_4}{j_5}{j_6}{j_7}{j_8}{j_9} &\equiv \sum_{m_1 ... m_9}
\threej{j_1}{m_1}{j_2}{m_2}{j_3}{m_3} \threej{j_4}{m_4}{j_5}{m_5}{j_6}{m_6} \notag \\
&\phantom{{}={}}\times \threej{j_7}{m_7}{j_8}{m_8}{j_9}{m_9} \threej{j_1}{m_1}{j_4}{m_4}{j_7}{m_7} \notag \\
&\phantom{{}={}}\times  \threej{j_2}{m_2}{j_5}{m_5}{j_8}{m_8} \threej{j_3}{m_3}{j_6}{m_6}{j_9}{m_9} \, ,
\end{align}
can be represented by the Yutsis graph given by the following hexagon
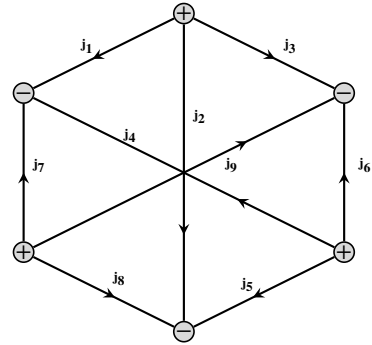
\diagram{9j_def}
involving six vertices and nine lines by inverting the signs of the magnetic quantum numbers in the last three $3jm$-symbols. While higher-order $3nj$-symbols only rarely arise in nuclear many-body theory, they can be equally represented by an unfactorizable Yutsis graph. They do in fact naturally enter in the partial-wave decomposition of nuclear $k$-body Hamiltonians for $k\geq4$.

In practice, Wigner $3nj$-symbols play an important role given that they can be pre-calculated and stored in cache in large-scale applications. This is typically done for $6j$-symbols and if necessary for (a subset of) $9j$-symbols. Since the number of $9j$-symbols is very large for a selected model space it is often useful to re-express $9j$-symbols as sums of products of $6j$-symbols and resort to much smaller $6j$-caches if the structure of the angular-momentum networks supports such a strategy.

\subsection{From tensor networks to Yutsis graphs}

The crucial first step consists of extracting the Yutsis graph associated with the SU-TN of interest. Following step (1) in Sec.~\ref{rationale}, this is achieved by expressing the original SU-T's in terms of SR-T's and a set of CG coefficients that are consecutively replaced by their $3jm$-symbol equivalents. The next step consists of splitting each involved summation according to
\begin{align}
\sum_{k} \quad \rightarrow \sum_{n_k l_k j_k t_k m_k } \rightarrow \quad \sum_{\tilde k} \sum_{m_k } \, .
\end{align}
In doing so, one can isolate the networks of $3jm$-symbols along with the sums over the magnetic quantum numbers. This corresponds to extracting the associated Yutsis graph.

\subsection{\label{sec:factorization}Factorization rules}

Having the Yutsis graph at hand, the goal is to simplify it a much as possible. This corresponds to identifying specific subparts in the graph that can be reduced via the application of identities satisfied by appropriate (sub)sets of $3jm$-symbols. Once this is completed, one is left with an expression involving irreducible Wigner $3nj$-symbols (see Sec.~\ref{unfactorizable}) and no magnetic quantum number dependence anymore.

The benefit of using Yutsis graphs is that the search for reducible parts can be automated while their actual reduction can be realized by applying systematic factorization rules on the graph. The rules are characterized by the length of the cycles involved in the factorization process. Below, the factorization rules are introduced one after another with increasing degree of complexity, i.e., cycle length. For the proofs of the factorization formula, the reader is referred to Ref.~\cite{Wormer2006}. A more extensive list of angular-momentum-algebra identities that can be used to define factorization rules can be found in Ref.~\cite{VaMo88}.

\subsection{Zero-line rule}

The most elementary simplification rule relates to the handling of a $3jm$-symbol with one vanishing $(j,m)$ pair, called \emph{zero line}. The corresponding vertex is represented as\footnote{The zero-line does not carry a direction given that the corresponding magnetic quantum number is neither positive nor negative.}
\diagram{zeroline}
and, with resort to Eq.~\eqref{eq:zero line}, could be reduced to a simple edge.

\subsubsection{Cycles of length two}
\begin{figure}
\centering
\includegrtex{bubble_rule}
\caption{Factorization rule for a 2-cycle giving rise to two Kronecker deltas plus a $3j$-symbol.}
\label{fig:bubblerule}
\end{figure}

The next simplest factorization corresponds to the reduction of a $2$-cycle. Algebraically, the corresponding identity is the \emph{orthogonality relation}
\begin{align}
\text{YG}_{\text{2c}} & \equiv  \sum_{m_1 m_2}  \threej{j_1}{m_1}{j_2}{m_2}{j_3}{m_3} \threej{j_{1}}{m_{1}}{j_2}{m_{2}}{j_{3^\pr}}{m_{3^\pr}}
\notag\\
&=  \sum_{m_1 m_2}  (-1)^{\sum_{k=1}^{3}(j_k -m_k)} \threej{j_1}{-m_1}{j_2}{-m_2}{j_3}{-m_3} \threej{j_{1}}{m_{1}}{j_2}{m_{2}}{j_{3^\pr}}{m_{3^\pr}}
\notag\\
&= \frac{1}{\hat{\jmath}_3^2} \delta_{j_3 j_{3^\pr}} \delta_{m_3 m_{3^\pr}} \trij{j_1}{j_2}{j_3}  \, .
\label{eq:threejOR}
\end{align}
Figure~\ref{fig:bubblerule} provides the diagrammatic representation of the identity stated in Eq.~\eqref{eq:threejOR}. Thus, the 2-cycle rule replaces two vertices connected by two lines by a single line in a Yutsis graph.
\subsubsection{Cycles of length three}

\begin{figure}
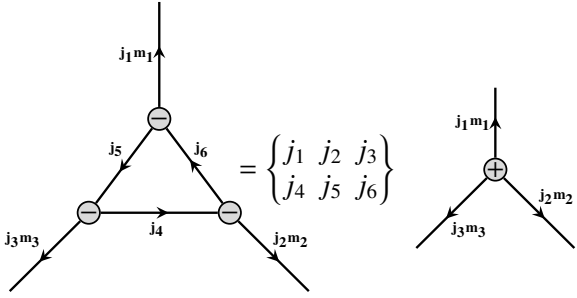

\centering
\scalebox{1.1}{\includegrtex{triangle_rule}}
\caption{Factorization rule for a 3-cycle giving rise to a single vertex plus a $6j$-symbol.}
\label{fig:trianglerule}
\end{figure}

The simplest factorization rule leading to a non-trivial Wigner $3nj$-symbol corresponds to the factorization of a 3-cycle as displayed in Fig.~\ref{fig:trianglerule}. Algebraically, the factorization corresponds to the identity
\begin{align}
\text{YG}_{\text{3c}} & \equiv    \sum_{m_4 m_5 m_6} (-1)^{\sum_{k=4}^{6} (j_k-m_k)}\threej{j_5}{m_5}{j_1}{m_1}{j_6}{-m_6} \notag \\
&\phantom{{}={}}\times
\threej{j_6}{m_6}{j_2}{m_2}{j_4}{-m_4}
\threej{j_4}{m_4}{j_3}{m_3}{j_5}{-m_5} \notag \\
&=
(-1)^{j_1+j_2+j_3} \threej{j_1}{m_1}{j_2}{m_2}{j_3}{m_3}
\sixj{j_1}{j_2}{j_3}{j_4}{j_5}{j_6} \, .
\label{eq:triangle}
\end{align}
Equation~\eqref{eq:triangle} allows one to factorize a topology involving three vertices into an irreducible part, i.e. the $6j$-symbol, and a single vertex. Therefore, the resulting graph contains two vertices and three lines less than the initial one.
\subsubsection{Cycles of length four}
\begin{figure*}
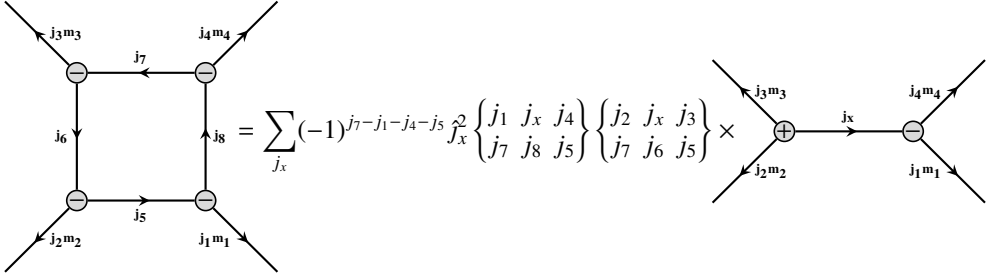

\centering
\includegrtex{quadrilateral_rule}
\caption{Factorization rule for a 4-cycle giving rise to a sum of terms with two $6j$-symbols multiplying a Yutsis graph containing two vertices.}
\label{fig:quadrilateralrule}
\end{figure*}
The most involved factorization rule employed in this work corresponds to a cycle of length four as displayed in Fig.~\ref{fig:quadrilateralrule}. The underlying algebraic identity is given by
\begin{align}
\text{YG}_{\text{4c}} & \equiv  \sum_{ \substack{ m_5 m_6 \\ m_7m_8 }}
(-1)^{j_5-m_5 +j_6 -m_6 +j_7 -m_7 +j_8-m_8}
\notag \\ &\phantom{{}={}}\times
\threej{j_8}{m_8}{j_1}{m_1}{j_5}{-m_5}
\threej{j_5}{m_5}{j_2}{m_2}{j_6}{-m_6}
\notag \\ &\phantom{{}={}}\times
\threej{j_6}{m_6}{j_3}{m_3}{j_7}{-m_7}
\threej{j_7}{m_7}{j_4}{m_4}{j_8}{-m_8}
\notag \\ &= (-1)^{j_7-j_1-j_4-j_5}
\sum_{j_x m_x}(-1)^{j_x-m_x} \hat{\jmath}_x^2
\notag \\ &\phantom{{}={}}\times
\threej{j_1}{m_1}{j_x}{-m_x}{j_4}{m_4}
\threej{j_2}{m_2}{j_x}{m_x}{j_3}{m_3}
\notag \\ &\phantom{{}={}}\times
\sixj{j_1}{j_x}{j_4}{j_7}{j_8}{j_5}
\sixj{j_2}{j_x}{j_3}{j_7}{j_6}{j_5} \, . \label{eq:square}
\end{align}
Equation~\eqref{eq:square} allows to factorize the topology involving four vertices into an irreducible part made of two $6j$-symbols and two vertices. Therefore, the resulting graph contains two vertices and three lines less than the initial one.

Note that the Yutsis graph in Fig.~\ref{fig:quadrilateralrule} has the symmetry of a square: rotations by multiples of $90^\circ$ leave it invariant. The rotation is equivalent to relabeling the edges, and leads to a different equivalent factorization for rotation angles of $90^\circ$ and $270^\circ$. The handling of cycles of length four is thus nonunique.

\subsubsection{Cycles beyond length four}

\begin{figure}
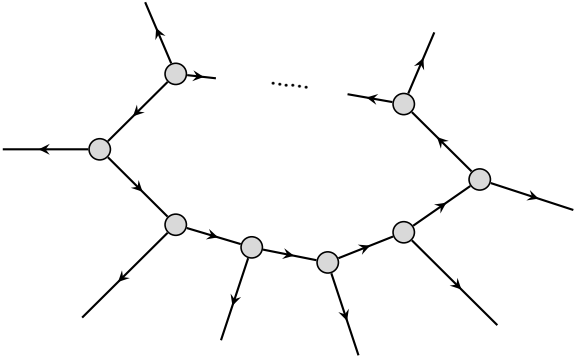

\centering
\includegrtex{higherorder_rule}
\caption{Schematic picture of a generic higher-order topology that cannot be simplified in terms of the triangle or the quadrilateral rules. Vertex signs are left out for simplicity.}
\label{fig:higherorderrule}
\end{figure}

While the present code supports factorizations involving up to cycles of length four, there exist topologies, sketched in Fig.~\ref{fig:higherorderrule}, which cannot be simplified through the above stated rules but require more involved identities. In principle, this restricts the range of applicability to topologies that do not contain cycles of length five or higher. The smallest cubic graph involving a cycle of length five is the so-called \emph{Peterson graph} containing ten vertices and 15 edges. Consequently, the simplest many-body diagram potentially leading to this topology must contain at least five two-body vertices, e.g., corresponding to a fifth-order MBPT diagram or a CC diagram with $T_3$ amplitudes.

In the testing phase of the current version of the code, the factorization rules were applied to hundreds of many-body diagrams including topologies that are far beyond current state-of-the-art applications. In none of these test cases a Yutsis graph involving a cycle of length five or higher appeared. Future versions of the program will be extended along these lines by including factorizations of more complex topologies, or including more elaborate techniques such as the interchange rule~\cite{vandyck03}.

%

\section{Applications}
\label{sec:appli}

A number of different many-body formalisms are now used to exemplify the steps at play in the symmetry reduction process of SU-TNs. The emphasis is on the angular-momentum reduction and details of the formalisms themselves are not within the scope of the present work. All considered examples are typical of state-of-the-art nuclear structure applications.

The formulae derived below are not in the computationally most optimized form. For instance, below parity- and isospin-conservation can be further exploited to yield more efficient implementations. However, processing $SU(2)$-symmetry yields by far the highest computational benefit due to larger dimensionality of the associated IRREPs.

\subsection{Many-body perturbation theory}

In many-body perturbation theory (MBPT) an infinite power series is taken as an ansatz for the exact ground-state energy and wave function~\cite{Shavitt2009}
\begin{subequations}
\begin{align}
E_k(\lambda) &= E_k^{(0)} + \lambda E_k^{(1)} +  \lambda^2 E_k^{(2)} + ... \, ,\\
\ket{\Psi_k(\lambda)} &= \ket{\Phi } + \lambda \ket{\Psi_k^{(1)} } +  \lambda^2  \ket{\Psi_k^{(2)} } + ... \, ,
\end{align}%
\label{eq:mbpt}%
\end{subequations}%
where the lower index $k$ enumerates excited states in the spectrum and $\ket{\Phi } \equiv \ket{ \Psi_k^{(0)} }$ denotes the unperturbed reference state. The expansions in Eq.~\eqref{eq:mbpt} are evaluated at $\lambda=1$ to obtain the quantities corresponding to the original problem of interet. Since the following is exclusively concerned with the description of nuclear ground states, i.e., $k=0$, the subscript is dropped for simplicity.

The starting point is given by the definition of a splitting of the full Hamiltonian
\begin{align}
H = H_0 + \lambda H_1\, ,
\end{align}
into an unperturbed part $H_0$ and a perturbation $H_1$ such that the reference energy is given by
\begin{align}
E_\text{ref} \equiv \braket{ \Phi | H | \Phi } = E_0^{(0)} + E_0^{(1)}\, .
\end{align}
The first contribution to the \emph{correlation energy}
\begin{align}
\Delta E \equiv E_0 - E_\text{ref} \label{correlE}
\end{align}
is, thus, obtained at second order. The simplest choice is to take $\ket{ \Phi }$ as a Slater determinant, typically obtained as the solution of a $SU(2)$-restricted Hartree-Fock (HF) calculation. In recent years, more sophisticated vacua have been used in order to account for so-called static correlations in open-shell systems. Both multi-configurational reference states obtained from a configuration interaction (CI) diagonalization in a small model space~\cite{Tichai:2018ncsmpt} and particle-number-broken HFB vacua~\cite{Tichai:2018mll} have shown to provide computationally cheap benchmarks without loss in accuracy when employing softened chiral potentials. For a recent review, see Ref.~\cite{Tichai:2020dna}.

Presently, the simplest single-reference case of low-order canonical HF-MBPT is discussed\footnote{Even though $\Delta E$ defined in Eq.~\eqref{correlE} is usually referred to as correlation energy it does not mean that $E_\text{ref}$ does not contain correlations effects, e.g., when using a symmetry-broken or multi-configurational vacuum. In the case of a HF vacuum, however, there is indeed no correlations contained in $\ket{\Phi }$ beyond those associated with Pauli's exclusion principle.}. Examples are worked out in detail to enable a deeper understanding of each of the individual algorithmic steps.

\subsubsection{Second-order energy correction}
\label{sec:mbpt2}

The second-order energy correction reads as\footnote{In this section a canonical HF vacuum is assumed throughout all derivations. More general choices give rise to one additional diagram at second order and eight additional diagrams at third order~\cite{Shavitt2009}. In any case the computational complexity is always driven by the canonical diagrams included in this discussion.}
\begin{align}
E_0^{(2)} = -\frac{1}{4} \sum_{abij} \frac{H_{abij} H_{ijab}}{\epsilon_{ij}^{ab} }\, ,
\label{eq:mbpt2}
\end{align}
where $a,b$ and $i,j$ denote particle and hole states, respectively, i.e., states that are unoccupied and occupied in the reference Slater determinant, respectively. Additionally, a short-hand notation for the energy denominator is used
\begin{align}
\epsilon_{ij...}^{ab...} \equiv \epsilon_a + \epsilon_b +... - \epsilon_i - \epsilon_j - ...\, ,
\end{align}
where $\epsilon_k$ denotes HF single-particle energies. According to the previous definitions Eq.~\eqref{eq:mbpt2} provides a closed SU-TN involving two mode-4 tensors, i.e. $H_{ijab}$ and $\epsilon_{ij}^{ab}$.

Expressing the two involved tensors in Eq.~\eqref{eq:mbpt2} in terms of their AMR-T counterparts according to (the inverse of) Eq.~\eqref{eq:twobodyJ} yields
\begin{align}
E_0^{(2)} &= -\frac{1}{4} \sum_{\tilde a \tilde b\tilde i\tilde j} \frac{1}{{\epsilon_{\tilde i\tilde j}^{\tilde a\tilde b} }} \sum_{\substack{J_1J_2 \\ M_1M_2}}  H^{J_1}_{\tilde a\tilde b\tilde i \tilde j} H^{J_2}_{\tilde i\tilde j \tilde a\tilde b}
\sum_{\substack{m_a m_b \\ m_i m_j }}
\clebsch{j_a}{m_a}{j_b}{m_b}{J_1}{M_1}
\notag \\ &\phantom{{}={}}\times
\clebsch{j_i}{m_i}{j_j}{m_j}{J_1}{M_1}
\clebsch{j_i}{m_i}{j_j}{m_j}{J_2}{M_2}
\clebsch{j_a}{m_a}{j_b}{m_b}{J_2}{M_2} \, ,
\label{eq:mbpt2cg}
\end{align}
where the fact was used that $\epsilon_{ij}^{ab} = \epsilon_{\tilde i\tilde j}^{\tilde a\tilde b}$ is already an AMR-T given that the single-particle energies are $m$-independent, i.e., $\epsilon_{\tilde k} = \epsilon_k$. The tensor network in Eq.~\eqref{eq:mbpt2cg} is split into an $SU(2)$-invariant part that does not depend on single-particle angular-momentum projection quantum numbers and a part carrying the full dependence of  magnetic quantum numbers that will be subsequently simplified. In a first step, CG coefficients are converted into $3jm$-symbols yielding
\begin{align}
E_0^{(2)} &= -\frac{1}{4} \sum_{\tilde a \tilde b\tilde i\tilde j} \frac{1}{{\epsilon_{\tilde i\tilde j}^{\tilde a\tilde b} }} \sum_{J_1J_2 }   H^{J_1}_{\tilde a\tilde b\tilde i \tilde j} H^{J_2}_{\tilde i\tilde j \tilde a\tilde b} \jhat{J}_1^2 \jhat{J}_2^2
\notag \\ &\times
\sum_{ M_1M_2} \sum_{\substack{m_a m_b \\ m_i m_j }} (-1)^{-2j_a+2j_b - 2M_1} (-1)^{-2j_i+2j_j - 2M_2}
\notag \\ &\phantom{{}={}}\times
\threej{j_a}{m_a}{j_b}{m_b}{J_1}{-M_1}
\threej{j_i}{m_i}{j_j}{m_j}{J_1}{-M_1}
\notag \\ &\phantom{{}={}}\times
\threej{j_i}{m_i}{j_j}{m_j}{J_2}{-M_2}
\threej{j_a}{m_a}{j_b}{m_b}{J_2}{-M_2}\, ,
\label{eq:mbpt2threej}
\end{align}
where each phase factor gives in fact
\begin{align}
(-1)^{-2j_a+2j_b - 2M_1} &= (-1)^{-2j_a+2j_b - 2(m_a + m_b)} \notag \\
 &= (-1)^{2(j_a - m_a) } (-1)^{2(j_b - m_b)} \notag \\
 & =1 \, .
\end{align}
Focusing on the $3jm$-symbols network in Eq.~\eqref{eq:mbpt2threej}, the second step consists of reversing all $m$ quantum numbers in the second and fourth $3jm$-symbols
\begin{align}
& \sum_{ M_1M_2} \sum_{\substack{m_a m_b \\ m_i m_j }} (-1)^{j_a-m_a+j_b-m_b+j_i-m_i+j_j-m_j+J_1-M_1+J_2-M_2}
\notag \\ &\phantom{{}={}}\times
\threej{j_a}{m_a}{j_b}{m_b}{J_1}{-M_1}
\threej{j_i}{-m_i}{j_j}{-m_j}{J_1}{M_1}
\notag \\ &\phantom{{}={}}\times
\threej{j_i}{m_i}{j_j}{m_j}{J_2}{-M_2}
\threej{j_a}{-m_a}{j_b}{-m_b}{J_2}{M_2}\, , \label{YGMBPT2}
\end{align}
at the price of an extra phase factor, where the magnetic quantum numbers have been added to the phase at no cost given that $m_i+m_j-M_1=0$ and $m_a+m_b-M_2=0$ hold and that $M_1$ and $M_2$ are integers. The expression in Eq.~\eqref{YGMBPT2} is now in the proper form to allow for its identification with an appropriate Yutsis graph
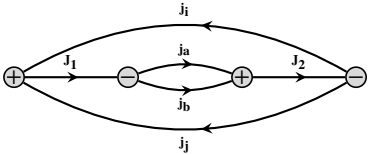
\diagram{mbpt2}
Now that the working graph has been built, the next step consists in simplifying it via the application of appropriate factorization rules. The application of the $2$-cycle rule, see Fig.~\ref{fig:bubblerule}, requires the direction of the edges carrying $j_a$ and $j_b$ to be reversed, thus bringing the phase $\Phi_\text{lr}=(-1)^{2j_a}(-1)^{2j_b}=(-1)^2=1$
and yielding the diagram
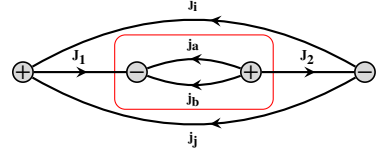
\diagram{mbpt2_step2}
where the red box indicates the subpart of the diagram that is factorizated in the next step.
Factorizing the $2$-cycle provides the intermediate factor
\begin{align}
  \frac{1}{\jhat{J}_1^{2}} \trij{j_a}{j_b}{J_1} \delta_{J_1J_2}
\end{align}
and leaves the diagram
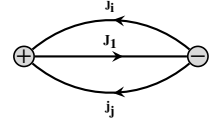
\diagram{mbpt2_step3}
In the last step, the $3j$-symbol is identified after reversing the orientation of the edges carrying $j_i$ and $j_j$
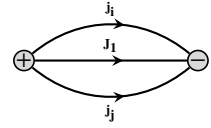
\diagram{mbpt2_step4}
leading to the additional phase $\Phi_\text{lr}=(-1)^{2j_i+2j_j}=(-1)^2=1$ and providing the overall result
\begin{align}
  \label{eq:mbpt2finalsign}
  \frac{1}{\jhat{J}_1^{2}} \trij{j_a}{j_b}{J_1} \trij{j_i}{j_j}{J_1} \delta_{J_1J_2} \, .
\end{align}
Replacing the $m$-dependent part of Eq.~\eqref{eq:mbpt2threej} by Eq.~\eqref{eq:mbpt2finalsign} finally provides the AMR form of the second-order energy correction
\begin{align}
  \label{eq:mbpt2sph}
  E_0^{(2)} = -\frac{1}{4} \sum_J \jhat{J}^2 \sum_{\tilde a\tilde b\tilde i\tilde j} \frac{H^J_{\tilde a\tilde b\tilde i\tilde j} H^J_{\tilde i\tilde j\tilde a\tilde b}}{\epsilon^{\tilde{a}\tilde{b}}_{\tilde{i}\tilde{j}} }\, ,
\end{align}
where triangular inequalities coming from $3j$-symbols are assumed. While the initial SU-TN is of $N^4$ complexity, the AMR-TN is of $\tilde N^4 \cdot (J_\text{max}+1)$ complexity, where $\tilde N$ is the number of reduced basis states $\tilde k$ and $J_\text{max}$ corresponds to the maximum number of channels (i.e. allowed values) of the two-body angular-momentum given the maximum one-body angular momentum retained in the (truncated) basis ${\cal B}_1$. For large model spaces the difference in runtime is improved by several orders of magnitude even for this very simple example.

\subsubsection{Third-order energy correction}

A more elaborate example is given by the third-order energy correction to the ground-state binding energy
\begin{align}
E_0^{(3)} \equiv E^{(3)}_{pp}+ E^{(3)}_{hh} + E^{(3)}_{ph} \, ,
\end{align}
which is divided into three contributions~\cite{Shavitt2009}
\begin{subequations}
\begin{align}
E^{(3)}_{pp} &= \frac{1}{8} \sum_{abcdij} \frac{H_{ijab} H_{abcd} H_{cdij} }{\epsilon^{ab}_{ij} \epsilon^{cd}_{ij}} \, ,\\
E^{(3)}_{hh} &= \frac{1}{8} \sum_{abijkl} \frac{H_{ijab} H_{klij} H_{abkl} }{\epsilon^{ab}_{ij} \epsilon^{ab}_{kl}} \, ,\\
E^{(3)}_{ph} &= - \sum_{abcijk} \frac{H_{ijab} H_{kbic} H_{ackj} }{\epsilon^{ab}_{ij} \epsilon^{ac}_{kj}} \, .
\end{align}
\end{subequations}
Following the same procedure as for $E_0^{(2)}$, one obtains the Yutsis graph associated with the particle-particle contribution (i.e. $E^{(3)}_{pp}$) is given by
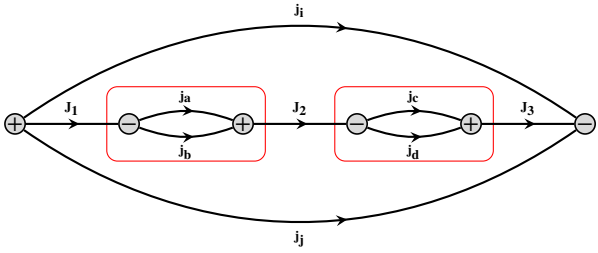
\diagram{mbpt3pp}
and, similarly, the one extracted from the hole-hole contribution (i.e. $E^{(3)}_{hh}$)
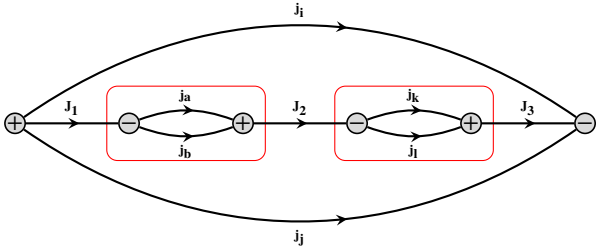
\diagram{mbpt3hh}
which are topologically identical. Due to the presence of one more Hamiltonian matrix element compared to the second-order energy correction, the number of $3jm$-symbols, i.e. the number of nodes, is increased by two. In both cases, the red boxes indicate the subgraphs that are factorized by the application of the $2$-cycle rule. Applying it twice and identifying the resulting Yutsis graph as a $3j$-symbol leads to the result
\begin{align}
  \frac{1}{\jhat{J}_1^{4}} \trij{j_a}{j_b}{J_1} \trij{j_c}{j_d}{J_1} \trij{j_i}{j_j}{J_1} \delta_{J_1J_2} \delta_{J_2J_3} \, .
\end{align}
Considering the $\jhat{J}_1^2 \jhat{J}_2^2 \jhat{J}_3^2$ factor coming from the prior conversion of CG coefficients into $3jm$-symbols, the final AMR form of the two contributions is
\begin{subequations}
  \begin{align}
    E_{pp}^{(3)}
    &= \frac{1}{8} \sum_J \jhat{J}^2 \sum_{\tilde{a}\tilde{b}\tilde{c}\tilde{d}\tilde{i}\tilde{j}}
    \frac{
      H^J_{\tilde{i} \tilde{j} \tilde{a} \tilde{b}}
      H^J_{\tilde{a} \tilde{b} \tilde{c} \tilde{d}}
      H^J_{\tilde{c} \tilde{d} \tilde{i} \tilde{j}}
    }{
      \epsilon^{\tilde{a}\tilde{b}}_{\tilde{i}\tilde{j}}
      \epsilon^{\tilde{c}\tilde{d}}_{\tilde{i}\tilde{j}}
    } \, , \\
    E_{hh}^{(3)}
    &= \frac{1}{8} \sum_J \jhat{J}^2 \sum_{\tilde{a}\tilde{b}\tilde{i}\tilde{j}\tilde{k}\tilde{l}}
    \frac{
      H^J_{\tilde{i} \tilde{j} \tilde{a} \tilde{b}}
      H^J_{\tilde{k} \tilde{l} \tilde{i} \tilde{j}}
      H^J_{\tilde{a} \tilde{b} \tilde{k} \tilde{l}}
    }{
      \epsilon^{\tilde{a}\tilde{b}}_{\tilde{i}\tilde{j}}
      \epsilon^{\tilde{a}\tilde{b}}_{\tilde{k}\tilde{l}}
    }\, ,
  \end{align}
\end{subequations}
which can be read as simple matrix-matrix products within each $J$ channel.

The symmetry reduction of the particle-hole term (i.e. $E^{(3)}_{ph}$) is more involved such that, following the same steps, the associated Yutsis graph is
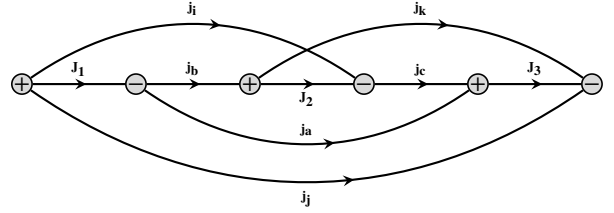
\diagram{mbpt3ph_naive}
and can be re-arranged in a more convenient way as
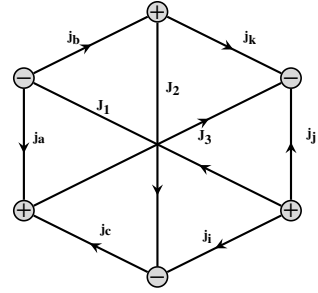
\diagram{mbpt3ph}
which is nothing but a $9j$-symbol. Consequently, the AMC form of the particle-hole term leads to the algebraic expression
\begin{align}
E_{ph}^{(3)} &= - \sum_K \jhat{K}^2 \sum_{J_1 J_2 J_3} \jhat{J}_1^2 \jhat{J}_2^2 \jhat{J}_3^2 \,
\ninej{J_1}{j_a}{j_j}{j_i}{J_2}{j_6}{j_b}{j_k}{J_3} \notag \\ &\phantom{{}={}}\times
\sum_{\tilde a\tilde b \tilde c\tilde i\tilde j \tilde k}
\frac{
H^{J_1}_{\tilde i\tilde j \tilde a\tilde b}
H^{J_2}_{\tilde k\tilde b\tilde i\tilde c}
H^{J_3}_{\tilde a\tilde c \tilde k\tilde j}
}{
\epsilon_{\tilde a\tilde b}^{\tilde i\tilde j}
\epsilon_{\tilde a\tilde c}^{\tilde k\tilde j}
}
\, . \label{3orderphAMR}
\end{align}
In practical applications, Eq.~\eqref{3orderphAMR} is conveniently re-written by expressing the $9j$-symbols as a sum of products of three $6j$-symbols
\begin{align}
E_{ph}^{(3)} &= - \sum_K \jhat{K}^2 \sum_{J_1 J_2 J_3} \jhat{J}_1^2 \jhat{J}_2^2 \jhat{J}_3^2 \notag \\ &\phantom{{}={}}\times
\sum_{\tilde a\tilde b \tilde c\tilde i\tilde j \tilde k}
\frac{
H^{J_1}_{\tilde i\tilde j \tilde a\tilde b}
H^{J_2}_{\tilde k\tilde b\tilde i\tilde c}
H^{J_3}_{\tilde a\tilde c \tilde k\tilde j}
}{
\epsilon_{\tilde a\tilde b}^{\tilde i\tilde j}
\epsilon_{\tilde a\tilde c}^{\tilde k\tilde j}
} \notag \\
&\phantom{{}={}}\times \sixj{j_i}{j_b}{J_1}{j_a}{j_j}{K}
\sixj{j_i}{j_b}{J_2}{j_k}{j_c}{K}
\sixj{j_k}{j_c}{J_3}{j_a}{j_j}{K}
\, , \label{3orderphAMRalternative}
\end{align}
which can be obtained graphically by a successive application of the $4$-cycle rule, the $3$-cycle rule and finally the identification of a redundant $3j$-symbol. Based on the introduction of so-called Pandya-transformed matrix elements~\cite{Su07}
\begin{align}
\breve{O}^{J_1}_{pqrs} \equiv - \sum_{J_2} \jhat{J}^2_2 \sixj{j_p}{j_q}{J_1}{j_r}{j_s}{J_2} O^{J_2}_{psrq} \, ,
\end{align}
Eq.~\eqref{3orderphAMRalternative} can eventually be written as
\begin{align}
E_{ph}^{(3)} = \sum_J \jhat{J}^2
\sum_{\tilde a\tilde b \tilde c\tilde i\tilde j \tilde k}
\frac{
\breve{H}^J_{\tilde i\tilde b \tilde a\tilde j}
\breve{H}^J_{\tilde a\tilde j \tilde k\tilde c}
\breve{H}^J_{\tilde k\tilde c \tilde i\tilde b}
}{
\epsilon_{\tilde a\tilde b}^{\tilde i\tilde j}
\epsilon_{\tilde a\tilde c}^{\tilde k\tilde j}
} \, ,
\label{eq:e3phpand}
\end{align}
which reads as the trace of a two-fold matrix-matrix product of Panya-transformed Hamiltonian matrix elements. Equation~\eqref{eq:e3phpand} clearly shows the computational benefit of an appropriate choice of the coupling order which in practice is not at all obvious.

\subsection{Coupled-cluster theory}

Contrary to a simple power series ansatz, coupled-cluster theory aims at a non-perturbative resummation of large classes of MBPT diagrams.

\subsubsection{General formalism}

 The starting point in the CC framework is an exponential ansatz to parameterize the exact ground state~\cite{Shavitt2009},
\begin{align}
\ket{\Psi } = e^{T} \ket{ \Phi } \, ,
\end{align}
in terms of the connected \emph{cluster operator} $T$ defined as
\begin{align}
T \equiv T_1 + T_2 + ... + T_A\, .
\label{eq:ccop}
\end{align}
The second-quantized form of the individual terms in Eq.~\ref{eq:ccop} is given by
\begin{align}
T_{n} \equiv \frac{1}{(n!)^2} \sum_{a_1...a_n} \sum_{i_1...i_n} t^{a_1 ...a _n}_{i_1 ...i_n} c^\dagger_{a_1} \cdots c^\dagger_{a_n} c_{i_n} \cdots c_{i_1} \, ,
\end{align}
with $t^{a_1 ...a _n}_{i_1 ...i_n}$ the $n$-tuple \emph{cluster amplitudes} characterizing a mode-$2n$ tensor. Thanks to the exponential form for the wave operator, the CC approach is manifestly size-extensize. In actual applications, $T$ is truncated at a fixed truncation level defining a particular CC model, e.g.,
\begin{subequations}
\begin{align}
T_\text{CCSD} &\equiv T_1 + T_2 \, \\
T_\text{CCSDT} &\equiv T_1 + T_2 + T_3 \, \\
&\shortvdotswithin{=} \notag
\end{align}
\end{subequations}
where the acronyms S,D,T,... indicate inclusion of single (S), double (D), triple (T), ... excitations.
Working equations can be conveniently be re-expressed in terms of the \emph{similarity-transformed Hamiltonian}
\begin{align}
\bar H &\equiv e^{-T} H e^T \notag \\
&= (He^T)_c \, ,
\end{align}
where the lower index $c$ stipulates the connected character of the expansion.

\subsubsection{Energy equation}

In the absence of three-body operators in the input Hamiltonian, the correlation energy is given for arbitrary CC truncations by
\begin{align}
\Delta E_\text{CC} &= \braket{ \Phi | \bar H | \Phi } \notag \\
&= \sum_{ai} t_{ai} f_{ia} + \sum_{abij} H_{abij} t_{ai} t_{bj}  + \frac{1}{4} \sum_{abij} H_{abij} t_{ijab}\, .
\label{eq:ECC}
\end{align}
Equation~\eqref{eq:ECC} defines a closed TN involving at most four internal contractions. Note that higher-order amplitudes affect the energy only implicitly by relaxing $T_1$ and $T_2$ without entering the energy equation explicitly.

Contrary to canonical MBPT, the CC energy equation involves mode-$2$ tensors associated with one-body operators, i.e. the $T_1$ amplitudes and the matrix elements $f_{pq}$ of the Fock operator. The Fock operator is $SU(2)$-invariant as long as the mean-field calculation is performed in a symmetry-restricted way. As the reference Slater determinant is presently characterized by $J=0$, cluster amplitudes are irreducible $SU(2)$ tensors of rank $J=0$ such that a similarity-transformed operator $\bar O$ has the same irreducible $SU(2)$ tensor rank as its non-transformed counterpart $O$. Hence, Wigner-Eckart's theorem trivially enables the introduction of reduced matrix elements
\begin{subequations}
\begin{align}
\braket{ \xi^\prime j^\prime m^\prime | T_1 | \xi j m }
&= \frac{1}{\jhat{\jmath}^\prime} \clebsch{j}{m}{0\,}{0\,}{j^\prime}{m^\prime}( \xi^\prime j^\prime  | \mathbf{T_{1}} | \xi j  ) \nonumber \\
&= (-1)^{j^\prime - m^\prime} \threej{j^\prime}{-m^\prime}{0}{0}{j}{m} ( \xi^\prime j^\prime  | \mathbf{T_{1}} | \xi j  ) \, ,  \\
\braket{ \xi^\prime j^\prime m^\prime | F | \xi j m } &= \frac{1}{\jhat{\jmath}^\prime} \clebsch{j}{m}{0\,}{0\,}{j^\prime}{m^\prime}( \xi^\prime j^\prime  | \mathbf{F} | \xi j  )\nonumber \\
&=(-1)^{j^\prime - m^\prime} \threej{j^\prime}{-m^\prime}{0}{0}{j}{m} ( \xi^\prime j^\prime  | \mathbf{F} | \xi j  ) \, .
\end{align}
\label{eq:WEt1f}
\end{subequations}
Inserting such forms into the first contribution to the CC the energy yields
\begin{align}
  \sum_{ai} f_{ia} t_{ai}
  &=
  \sum_{\xi_a \xi_i}
  \sum_{j_a j_i}
  ( \xi_i j_i \vert \mathbf{F}   \vert \xi_a j_a )
  ( \xi_a j_a \vert \mathbf{T_1} \vert \xi_i j_i ) \notag \\*
  &\hphantom{\,} \times
  \sum_{m_a m_i}
  (-1)^{j_i - m_i + j_a - m_a}
  \threej{j_i}{m_i}{0}{0}{j_a}{-m_a}
  \threej{j_a}{m_a}{0}{0}{j_i}{-m_i} \, ,
\end{align}
from which the $m$-dependent part can be extracted to yield the Yutsis graph
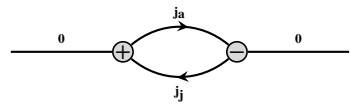
\diagram{ECC1}
Reversing the direction of the $j_a$ edge ($\Phi_\text{lr}=(-1)^{2j_a}=-1$) together with changing the sign of the leftmost node ($\Phi_\text{ns}=(-1)^{j_a+j_j}$) allows to make use of the $2$-cycle rule which leads to
\begin{align}
  - (-1)^{j_a+j_j} \trij{j_a}{j_j}{0} = - (-1)^{j_a+j_j} \delta_{j_a j_j} = \delta_{j_a j_j} \, ,
\end{align}
such that one obtains the final AMR form
\begin{align}
  \sum_{ai} f_{ia} t_{ai}
  &=
  \sum_{\xi_a \xi_i}
  \sum_{j_a}
  ( \xi_i j_a \vert \mathbf{F}   \vert \xi_a j_a  )
  ( \xi_a j_a \vert \mathbf{T_1} \vert \xi_i j_a  ) \, .
\end{align}
For the second term of the energy equation, one has
\begin{align}
  \label{eq:ecc2_start_coupling}
  \sum_{abij} H_{ijab} t_{ai} t_{bj}
  &=
  \sum_{\substack{\xi_a \xi_b \\ \xi_i \xi_j }}
  \sum_{\substack{j_a j_b \\ j_i j_j }}
  \sum_{J}
  \jhat{J}^2
  \, H^J_{\tilde i \tilde j \tilde a \tilde b}\notag \\*
  &\hphantom{=} \times
  \, ( \xi_a j_a \vert \mathbf{T_1} \vert \xi_i j_i )
  \, ( \xi_b j_b \vert \mathbf{T_1} \vert \xi_j j_j ) \notag \\*
  &\hphantom{=} \times
  \sum_{\substack{ m_a m_b \\ m_i m_j }} \sum_{M}
     \notag \\*
  &\hphantom{=\times } \times (-1)^{(j_i - j_j + M)+(j_a - j_b + M)}  \notag \\*
  &\hphantom{=\times } \times
  \threej{j_i}{m_i}{j_j}{m_j}{J}{-M}
  \threej{j_a}{m_a}{j_b}{m_b}{J}{-M} \notag \\*
  &\hphantom{=\times } \times (-1)^{(j_a-m_a)+(j_b-m_b)}\notag \\*
  &\hphantom{=\times } \times
  \threej{j_a}{m_a}{0}{0}{j_i}{-m_i}
  \threej{j_b}{m_b}{0}{0}{j_j}{-m_j} \, .
\end{align}
Reversing the signs of $m$ quantum numbers in the second $3jm$-symbol, the $m$-dependent part of Eq.~\eqref{eq:ecc2_start_coupling} delivers the Yutsis graph
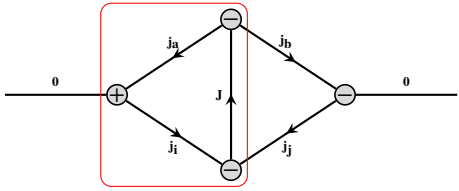
\diagram{ECC2}
with two external edges carrying zero angular momentum.
 Applying twice the zero-line rule, one ends up with the graphical representation of a $3j$-symbol, such that the $m$-dependent part of Eq.~\eqref{eq:ecc2_start_coupling} reduces to
\begin{align}
  \frac{1}{\jhat{j}_a \jhat{j}_b} \delta_{j_a j_i} \delta_{j_b j_j} \trij{j_a}{j_b}{J} \, ,
\end{align}
thus providing the final closed AMR-TN under the form\begin{align}
  \sum_{abij} H_{ijab} t_{ai} t_{bj}
  &=
  \sum_{\substack{\xi_a \xi_b \\ \xi_i \xi_j }}
  \sum_{j_a j_b J}
  \frac{\jhat{J}^2}{\jhat{j}_a \jhat{j}_b}
  H^J_{\xi_i j_a \xi_j j_b \tilde a \tilde b}\notag \\*
  &\hphantom{=} \times
  \, ( \xi_a j_a \vert \mathbf{T_1} \vert \xi_i j_a )
  \, ( \xi_b j_b \vert \mathbf{T_1} \vert \xi_j j_b ) \, .
\end{align}
The detailed derivation of the last contribution to the energy equation is omitted given that it is formally identical to the derivation of the second-order MBPT correction, i.e. the appropriate Yutsis graph is the one displayed in Sec.~\ref{sec:mbpt2}.
The final result reads as
\begin{align}
  \label{eq:ECC3}
  \frac{1}{4} \sum_{abij} H_{ijab} t_{abij}
  &= \frac{1}{4} \sum_J \jhat{J}^2 \sum_{\tilde a\tilde b\tilde i\tilde j} H^J_{\tilde i\tilde j\tilde a\tilde b} \, t^J_{\tilde a\tilde b\tilde i\tilde j} \, .
\end{align}

\subsubsection{Amplitude equations}

The unknown cluster amplitudes are obtained by solving a set of CC \emph{amplitude equations}
\begin{subequations}
\begin{align}
0 &= \braket{ \Phi^a_i | \bar H | \Phi } \, , \\
0 &= \braket{ \Phi^{ab}_{ij} | \bar H | \Phi } \, , \\
&\shortvdotswithin{=} \notag
\end{align}
\label{eq:CCamp}
\end{subequations}
Equations~\eqref{eq:CCamp} constitute a set of coupled non-linear equations that must be solved iteratively for every external index combination. They also provide typical examples of open SU-TNs containing external indices that are not summed over\footnote{In Ref.~\cite{Tichai:2019ksh}, the AMR of the triple BCC amplitudes evaluated at second order in perturbation theory was performed. It constituted the first application of the presently introduced numerical tool. The analytical expressions of the open AMR-TN's corresponding to the 20 different contributions were provided as a testimony of the complexity at play.}.

In order to perform the symmetry reduction, one must sum over all magnetic quantum numbers, and in particular the external ones.
This will lead to a closed Yutsis graph.
To do so, an external coupling order has to be fixed.
The coupling
\begin{align}
  \frac{1}{\jhat{j}_a^2} \sum_{m_a m_i} \clebsch{j_a}{m_a}{\,0\,}{\,0\,}{j_i}{m_i}
\end{align}
is used in the case of the $T_1$ amplitude equations and is such that
\begin{align}
  \frac{1}{\jhat{j}_a^2} \sum_{m_a m_i} \clebsch{j_a}{m_a}{\,0\,}{\,0\,}{j_i}{m_i} t_{ai}
  &= \frac{1}{\jhat{j}_a} ( \xi_{a} j_{a} \vert \mathbf{T_1} \vert \xi_{i} j_{a} ) \, ,
\end{align}
whereas the coupling
\begin{align}
  \label{eq:t2_external_coupling}
  \frac{1}{\jhat{J}^2} \sum_{m_a m_b m_i m_j M} \clebsch{j_a}{m_a}{j_b}{m_b}{J}{M} \clebsch{j_i}{m_i}{j_j}{m_j}{J}{M}
\end{align}
is used in the case of the $T_2$ amplitude equations and is such that
\begin{align}
  \frac{1}{\jhat{J}^2} \sum_{m_a m_b m_i m_j M} \clebsch{j_a}{m_a}{j_b}{m_b}{J}{M} \clebsch{j_i}{m_i}{j_j}{m_j}{J}{M} t_{abij}
  &= t^J_{\tilde{a}\tilde{b}\tilde{i}\tilde{j}} \, .
\end{align}
The alternative couplings\footnote{Such a choice is sometimes referred to as \textit{cross coupling} since it involves angular-momentum coupling of bra and ket single-particle states.}
\begin{align}
  \frac{1}{\jhat{J}^2} \sum_{m_a m_b m_i m_j M} \clebsch{j_a}{m_a}{j_i}{m_i}{J}{M} \clebsch{j_b}{m_b}{j_j}{m_j}{J}{M}
\end{align}
or even
\begin{align}
  \frac{1}{\jhat{J}^2} \sum_{m_a m_b m_i m_j M} \clebsch{j_a}{m_a}{j_j}{m_j}{J}{M} \clebsch{j_b}{m_b}{j_i}{m_i}{J}{M}
\end{align}
can be used equally well. However, it turns out that the resulting equation will be much simpler when the first option is employed since the coupling order is consistent with the coupling order used for the Hamiltonian matrix elements. This is an example where prior experience provides a strong guidance for the proper choice of the angular-momentum coupling scheme even though ultimately all choices yield equivalent results.

To exemplify the coupling of open SU-TNs, one particular contribution to the CCSD doubles amplitude equation is chosen
\begin{align}
D_{abij}&\equiv\sum_{kl} \sum_{ cd} H_{klcd} t_{dj} t_{ak} t_{cbil} \, , \label{eq:ccsd_double_term}
\end{align}
where ($k,l,c,d$) denote internal indices that are summed over while ($a,b,i,j$) characterize the external indices.
The construction of the angular-momentum network originating from the application of the external coupling, defined in Eq.~\eqref{eq:t2_external_coupling}, to Eq.~\eqref{eq:ccsd_double_term} requires to sum over a product of
\begin{enumerate}
  \item[$(i)$] two $3jm$-symbols coming from the external coupling of $a,b$ and $i,j$,
  \item[$(ii)$] two $3jm$-symbols with zero-edges coming from the application of Wigner-Eckart theorem to the $T_1$ amplitudes,
  \item[$(iii)$] four $3jm$-symbols coming from the coupling of $H$ and $T_2$ matrix elements,
\end{enumerate}
yielding eight $3jm$-symbols and eleven summations over magnetic quantum numbers, eight corresponding to one-body indices ($m_a,m_b,m_c,m_d,m_i,m_j,m_k,m_l$), two originating from the decoupling of $H$ and $T_2$ ($M_1$,$M_2$) and one ($M$) coming from the external coupling of double amplitude equation.
The corresponding Yutsis graph is given by
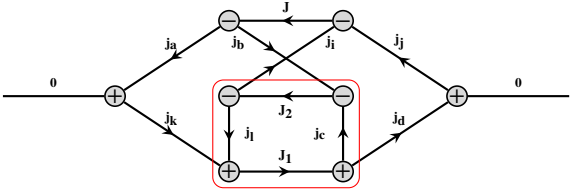
\diagram{CCSDamp}
The red box indicates a subgraph to which the 4-cycle factorization rule can be applied. However, first applying twice the zero-line rule to the leftmost and rightmost nodes ($\frac{1}{\hat{j}_a} \delta_{j_a j_k}$ and $\frac{1}{\hat{j}_j} \delta_{j_j j_d}$) directly yields a Yutsis graph that is topologically equivalent to the one of the third-order particle-hole contribution in MBPT, i.e., which corresponds to a $9j$-symbol. The final expression reads as
\begin{align}
D^J_{\tilde{a}\tilde{b}\tilde{i}\tilde{j}} & = \sum_{J_1 J_2 K} \frac{\jhat{J}^2_1 \jhat{J}^2_2 \jhat{K}^2}{\jhat{\jmath}_a \jhat{\jmath}_j} \sum_{\xi_{k}\xi_{d}\tilde{l}\tilde{c}} H^{J_1}_{\xi_{k}j_a\tilde{l}\tilde{c}\xi_{d}j_j}\notag \\&{} \times
  ( \xi_{d} j_{j} \vert \mathbf{T_1} \vert \xi_{j} j_{j} )
  ( \xi_{a} j_{a} \vert \mathbf{T_1} \vert \xi_{k} j_{a} )
  \, t^{J_2}_{\tilde{c}\tilde{b}\tilde{i}\tilde{l}}
\notag \\&{} \times
\sixj{j_i}{j_b}{K}{j_a}{j_j}{J}
\sixj{j_c}{j_l}{K}{j_a}{j_j}{J_1}
\sixj{j_i}{j_l}{J_2}{j_b}{j_c}{K}.
\end{align}

\subsection{In-medium similarity renormalization group}

As a final example, the IMSRG approach is considered providing a non-perturbative alternative to CC theory. Throughout the last decade, IMSRG has been successfully applied to various nuclear observables, including low-lying excited states and electromagnetic transitions whose treatments were pioneered and applied to mid-mass closed-shell nuclei in Ref.~\cite{Parzuchowski:2017wcq}.  Without the use of angular-momentum reduction, such studies in the mid-mass regime would have been impossible from a computational point of view. Thus, non-scalar operators associated with, e.g., electromagnetic transitions constitute an excellent playground to yet extend the application of our automated treatment of angular-momentum reduction.

\subsubsection{General formalism}

The IMSRG formalism is based on a unitary transformation $U(s)$ of operators parametrized by a continuous variable $s\in [0,\infty)$ such that
\begin{align}
O(s) = U(s) O(0) U^\dagger(s) \, \label{IMSRGtransfo} .
\end{align}
Equation~\eqref{IMSRGtransfo} can be recast into a first-order ordinary differential equation (ODE)
\begin{align}
\frac{\text{d}}{\text{d}s} O(s) = [ \eta(s), O(s) ] \, ,
\label{eq:imsrgcomm}
\end{align}
involving an anti-Hermitian generator $\eta$ that can be chosen conveniently to obtain a desired decoupling pattern. A standard choice is given by the \textit{Wegner generator}
\begin{align}
\eta(s) = [H_\text{od}(s), H_\text{d}(s)] \, ,
\end{align}
defined as the commutator of the suitably chosen 'diagonal' and 'off-diagonal' parts of $H$, the end result being that $H_\text{od}(s)$ is eventually driven to zero. Even though the initial operator may contain up to two-body parts only, the evaluation of the commutator in Eq.~\eqref{eq:imsrgcomm} increases the particle rank of the operator, thus, inducing many-body operators up to the $A$-body level. In practice,  the IMSRG(2) truncation is typically employed in which operators of higher rank than two-body operators are discarded. As discussed in Ref.~\cite{Hergert:2015awm}, the IMSRG(2) approximation is exact up to third order in MBPT for the ground-state energy while resumming large classes of higher-order diagrams.

\subsubsection{Evolution of non-scalar operators}

The form to Eq.~\eqref{eq:imsrgcomm} is completely generic and valid for an arbitrary Hermitian operator $O$, independently of its transformation properties with respect to $SU(2)$ symmetry. For practical applications,  the specific tensorial properties of $O$ need however to be taken into account. The evaluation of the ground-state energy provides the simplest case since both $O=H$ and the generator are scalar operators in this case.

In the general case where $O$ is a spherical tensor operator of rank $\lambda$, the evaluation of the AMR form of the commutator appearing in Eq.~\eqref{eq:imsrgcomm} is key. The associated form can be generically written as
\begin{align}
  C^\lambda_\mu
  &\equiv [ \mathbf{S}^{\lambda_1}, \mathbf{T}^{\lambda_2} ]^\lambda_\mu = [\mathbf{S}^{\lambda_1} \mathbf{T}^{\lambda_2}]^\lambda_\mu - [\mathbf{T}^{\lambda_2} \mathbf{S}^{\lambda_1}]^\lambda_\mu \, , \label{commAMRform}
\end{align}
where $S^{\lambda_1}_{\mu_1}$ and $T^{\lambda_2}_{\mu_2}$ are spherical tensor operator of rank $\lambda_1$ and $\lambda_2$, respectively, which are subsequently coupled to give a tensor of rank $\lambda$. This coupling is obtained via \emph{spherical tensor product} defined through
\begin{align}
  \label{eq:tensprod}
  [\mathbf{S}^{\lambda_1} \mathbf{T}^{\lambda_2} ]^\lambda_\mu
  &\equiv \sum_{\mu_1 \mu_2} \clebsch{\lambda_1}{\mu_1}{\lambda_2}{\mu_2}{\,\lambda\,}{\,\mu\,}
  S^{\lambda_1}_{\mu_1} T^{\lambda_2}_{\mu_2} \, ,
\end{align}
where the left-hand-side is indeed a spherical tensor operator of rank $\lambda$.

While the complete list of contributions can be found in Ref.~\cite{Parzuchowski:2017wcq}, the so-called particle-particle contribution to the two-body part of the evolved operator is considered as an example. The associated SU-TN expression is given by
\begin{align}
  \label{eq:mschemetensex}
  C^{\lambda\mu}_{pqrs}
  &= \frac{1}{2} \sum_{tu} \bar{n}_t \bar{n}_u \sum_{\mu_1 \mu_2} \clebsch{\lambda_1}{\mu_1}{\lambda_2}{\mu_2}{\,\lambda\,}{\,\mu\,} S^{\lambda_1\mu_1}_{pqtu} T^{\lambda_2\mu_2}_{turs} \, ,
\end{align}
where $n_p \in \{0,1\}$ denotes the occupation number of the state $\ket{p}$ and $\bar{n}_p \equiv 1-n_p$. The occupation number is independent of the projection quantum number, i.e. $n_p=n_{\tilde p}$ as well as $\bar{n}_p=\bar{n}_{\tilde{p}}$.

Applying WET to the left-hand-side of Eq.~\eqref{eq:mschemetensex} provides
\begin{align}
C^{\lambda\mu}_{pqrs}  &= \sum_{J_1 J_2 M_1 M_2} \frac{1}{\jhat{J}_1}
\clebsch{j_p}{m_p}{j_q}{m_q}{J_1}{M_1}
\clebsch{j_r}{m_r}{j_s}{m_s}{J_2}{M_2} \notag \\ &\phantom{{}={}}\times
\clebsch{J_2}{M_2}{\lambda}{\mu}{J_1}{M_1}
(\tilde p \tilde q J_1 | \mathbf{C}^\lambda | \tilde r \tilde s J_2 ) \, .
\end{align}
and similarly for the tensors operators arising from commutator expansion
\begin{subequations}
\begin{align}
S^{\lambda_1 \mu_1}_{pqtu}  &= \sum_{J_3 J_4 M_3 M_4} \frac{1}{\jhat{J}_3}
\clebsch{j_p}{m_p}{j_q}{m_q}{J_3}{M_3}
\clebsch{j_t}{m_t}{j_u}{m_u}{J_4}{M_4} \notag \\ &\phantom{{}={}}\times
\clebsch{J_4}{M_4}{\lambda_1}{\mu_1}{J_3}{M_3}
(\tilde p \tilde q J_3 | \mathbf{S}^{\lambda_1} | \tilde t \tilde u J_4 ) \, , \\
T^{\lambda_2 \mu_2}_{turs}  &= \sum_{J_5 J_6 M_5 M_6} \frac{1}{\jhat{J}_5}
\clebsch{j_t}{m_t}{j_u}{m_u}{J_5}{M_5}
\clebsch{j_r}{m_r}{j_s}{m_s}{J_6}{M_6} \notag \\ &\phantom{{}={}}\times
\clebsch{J_6}{M_6}{\lambda_2}{\mu_2}{J_5}{M_5}
(\tilde t \tilde u J_5 | \mathbf{T}^{\lambda_2} | \tilde r \tilde s J_6 ) \, .
\end{align}
\label{eq:STtens}
\end{subequations}
In the following the standard external coupling of a tensor operator (see Eq.~\eqref{eq:t2_external_coupling} for the scalar case) is employed
\begin{align}
\label{eq:tensor_external_coupling}
\frac{1}{\jhat{J}_1}\sum_{\substack{m_a m_b \\ m_i m_j }} \sum_{M_1 M_2 \mu} \clebsch{j_p}{m_p}{j_q}{m_q}{J_1}{M_1} \clebsch{j_r}{m_r}{j_s}{m_s}{J_2}{M_2} \clebsch{J_2}{M_2}{\lambda}{\mu}{J_1}{M_1}
\end{align}
Applying Eq.~\eqref{eq:tensor_external_coupling} to Eq.~\eqref{eq:mschemetensex} and inserting all the transformation displayed in Eqs.~\eqref{eq:STtens} yields
\begin{align}
\label{Jschmecomm}
\MoveEqLeft(\tilde p \tilde q J_1 | \mathbf{C}^\lambda | \tilde r \tilde s J_2 ) = \notag \\
&\frac{1}{2}\sum_{\mu_1 \mu_2 \mu} \sum_{\{m_i\}} \sum_{\substack{J_1,...,J_6 \\ M_1,...,M_6  }} \frac{1}{\jhat{J}_1 \jhat{J}_3 \jhat{J}_5 }
\clebsch{\lambda_1}{\mu_1}{\lambda_2}{\mu_2}{\,\lambda\,}{\,\mu\,} \notag \\ &\times
\clebsch{j_p}{m_p}{j_q}{m_q}{J_1}{M_1} \clebsch{j_r}{m_r}{j_s}{m_s}{J_2}{M_2} \clebsch{J_2}{M_2}{\lambda}{\mu}{J_1}{M_1}
\notag \\ &\times
\clebsch{j_p}{m_p}{j_q}{m_q}{J_3}{M_3}
\clebsch{j_t}{m_t}{j_u}{m_u}{J_4}{M_4}
\clebsch{J_4}{M_4}{\lambda_1}{\mu_1}{J_3}{M_3} \notag \\ &\times
\clebsch{j_t}{m_t}{j_u}{m_u}{J_5}{M_5}
\clebsch{j_r}{m_r}{j_s}{m_s}{J_6}{M_6}
\clebsch{J_6}{M_6}{\lambda_2}{\mu_2}{J_5}{M_5} \notag \\ &\times
\bar{n}_{\tilde t}\bar{n}_{\tilde u}
(\tilde p \tilde q J_3 | \mathbf{S}^{\lambda_1} | \tilde t \tilde u J_4 )
(\tilde t \tilde u J_5 | \mathbf{T}^{\lambda_2} | \tilde r \tilde s J_6 ) \, .
\end{align}
After transforming CG coefficients into $3jm$-symbols, the angular-momentum network appearing in Eq.~\eqref{Jschmecomm} can be identified with the following Yutsis graph
\diagram{IMSRGtens}
in which the red boxes indicate the subgraphs to which the 2-cycle factorization rule is applied.
The residual Yutsis graph corresponds to a $6j$-symbol and a phase factor, such that one eventually obtains the reduced for of Eq.~\eqref{Jschmecomm} as
\begin{align}
\MoveEqLeft(\tilde p \tilde q J_1 | \mathbf{C}^\lambda | \tilde r \tilde s J_2 ) \notag \\
&=
\frac{1}{2} \jhat{\lambda} (-1)^{J_1 + J_2 + \lambda} \sum_{J_3} \sixj{\lambda_1}{\lambda_2}{\lambda}{J_2}{J_1}{J_3}  \sum_{\tilde t \tilde u}
\notag \\ &\phantom{{}={}}\times
\bar{n}_{\tilde t}\bar{n}_{\tilde u}
(\tilde p \tilde q J_1 | \mathbf{S}^{\lambda_1} | \tilde t \tilde u J_3 )
(\tilde t \tilde u J_3 | \mathbf{T}^{\lambda_2} | \tilde r \tilde s J_2 ) \, .
\label{eq:imsrgtens}
\end{align}
In the special case of scalar operator, i.e., $\lambda = \lambda_1=\lambda_2=0$, the AMC form of the commutator simplifies to
\begin{align}
 \label{eq:imsrg_scalar_comm}
\MoveEqLeft(\tilde p \tilde q J | \mathbf{C}^\lambda | \tilde r \tilde s J )
\notag\\&
=
\frac{1}{2\jhat{J}}  \sum_{\tilde t \tilde u} \bar n_{\tilde t} \bar n_{\tilde u} (\tilde p \tilde qJ| \mathbf{S}^0| \tilde t \tilde u J) (\tilde t \tilde u J | \mathbf{T} ^0 | \tilde r \tilde s J) \, ,
\end{align}
where the following property of the $6j$-symbol
\begin{align}
\sixj{0}{0}{0}{j_1}{j_2}{j_3} = (-1)^{2j_1} \frac{1}{\jhat{j}_1} \delta_{j_1 j_2} \delta_{j_2 j_3}
\end{align}
has been used. Equation~\eqref{eq:imsrg_scalar_comm} can be rewritten in terms of angular-momentum-coupled matrix elements (Eq.~\eqref{eq:twobodyJ}) instead of reduced matrix elements giving
\begin{align}
C_{\tilde p \tilde q \tilde r \tilde s}^J = \frac{1}{2}  \sum_{\tilde t \tilde u} \bar n_{\tilde t} \bar n_{\tilde u}
S_{\tilde p \tilde q \tilde t \tilde u}^J T_{\tilde t \tilde u \tilde r \tilde s}^J \, .
\end{align}

\section{Conclusions}
\label{sec:conc}

In the present work, an automated tool to perform symbolic angular-momentum algebra operations has been designed. This tool relates to the fact that the working equations, i.e. the symmetry-unrestricted tensor networks, at play in state-of-the-art nuclear many-body methods can be analytically reduced with respect to angular-momentum quantum numbers whenever they are effectively employed in a symmetry-restricted context. The corresponding time-consuming and error-prone derivation of the angular-momentum-reduced form of the tensor networks is thus performed in a matter of seconds. The design of the tool is based on the use of Yutsis graph representing networks of Wigner $3jm$-symbols and fulfilling sets of factorization rules whose repeated application eventually provides the angular-momentum-reduced form of the equations. While examples of applications have been provided for many-body perturbation theory, coupled cluster theory and the in-medium similarity renormalization group method, the code can be interfaced with any many-body formalism of interest.

While the present paper focuses on $SU(2)$ symmetry, extensions are envisioned for the future, e.g. to the subgroup of $SU(2)$ at play in axially deformed nuclei, or to other symmetry groups.

In view of obtaining the error prone, fast and numerically optimized implementation of involved many-body formalisms, the present code serves as the missing link between an automated tool used to generate the initial symmetry-unrestricted equations and and an automated tool used to produce the efficient source code dedicated to numerical applications.

\section{Command-Line Interface and Input Files}

For simple usage of the code, the \amc{} program is provided. The \amc{} program is a command-line interface to the code that can be used to reduce a set of equations and output the reduced equations to a LaTeX document. The unreduced equations are supplied as an \AMC{} file, in a domain-specific language described in Sec.~\ref{sec:amclang}.

\subsection{Command-Line Options}

There are a few options, which can be passed to \amc{}, that modify the behavior of the program:
\begin{description}[font=\normalfont\ttfamily,style=nextline]
  \item[-o OUTPUT, -{-}output OUTPUT]
    Write the resulting LaTeX document to \textit{OUTPUT}.
    By default the code strips the extension from the input file, adds a \texttt{.tex} extension, and creates a file of that name in the same directory as the input file.
  \item[-{-}collect-ninejs]
    Try to reconstruct Wigner $9j$ symbols from products of $6j$ symbols in the reduced expressions.
    This results in shorter expressions, but might obscure opportunities to identify intermediates, e.g., when some of the $6j$ symbols only depend on the indices of single tensors.
  \item[-{-}keep-trideltas]
    Print triangular inequalities generated during the reduction process.
    Most often, these inequality constraints are implicitly contained in tensor variables, so removing them does not generate any loss of information.
    Constraints that are implicit in $6j$ or $9j$ symbols are never shown.
  \item[-{-}wet-convention CONVENTION]
    Switch the convention used for reduced matrix elements.
    Currently, the code supports two conventions: the \texttt{wigner} convention
    \begin{align}
      \MoveEqLeft\braket{ \xi_1 j_1 m_1 | T^{J}_{M} | \xi_2 j_2 m_2 } \notag\\
        &= (-1)^{2J} \frac{1}{\jhat{\jmath}_1} \clebsch{j_2}{m_2}{J}{M}{j_1}{m_1}
        ( \xi_1 j_1  | {\mathbf{T}^{J}} | \xi_2 j_2 ),
    \end{align}
    which has been adopted in the body of the paper, and the \texttt{sakurai} convention
    \begin{align}
      \MoveEqLeft\braket{ \xi_1 j_1 m_1 | T^{J}_{M} | \xi_2 j_2 m_2 } \notag\\
        &= \frac{1}{\jhat{\jmath}_2} \clebsch{j_2}{m_2}{J}{M}{j_1}{m_1}
        ( \xi_1 j_1  | {\mathbf{T}^{J}} | \xi_2 j_2  ).
    \end{align}
    The \texttt{wigner} convention is used by default.
\end{description}

\subsection{\label{sec:amclang}Angular-Momentum Coupling Language}

To facilitate the use of the code the \AMC{} language is provided to specifiy the tensors and their coupling schemes as well as to enter the equations to be coupled. The basic building blocks of the language are integers and fractions (\lstinline!i/j!), strings (\lstinline!"abc"!), booleans (\lstinline!true, false!), and tuples [\lstinline!(x1, x2, ...)!]. Comments are introduced by the pound sign (\#) and last until the end of the line.

The tensors and equations are defined in a plain text file consisting of tensor declaration and equation statements.
The statement
\begin{amclisting}
declare $tensor$ {
  $key$ = $value$,
  ...
}
\end{amclisting}
declares a tensor with properties specified by the key-value pairs inside the curly braces.
The following keys are accepted:
\begin{description}
  \item[mode] The number of indices of the tensor. The mode is either specified as an even integer, e.g.\ 2 for a one-body operator or 4 for a two-body operator, or as a tuple of two numbers \lstinline!(x,y)! to specify $x$ creator and $y$ annihilator indices.
  \item[scalar] A boolean indicating that the tensor is scalar (rank 0). The code exploits additional angular-mo\-men\-tum constraints for scalar tensors, and uses the unreduced matrix elements by default.
  \item[reduce] A boolean indicating that this scalar tensor uses Wigner-Eckart reduced matrix elements. This key is ignored for nonscalar tensors, which always use reduced matrix elements. The default values is \lstinline!false!, so unreduced matrix elements are assumed.
  \item[diagonal] A boolean value that specifies whether the tensor is diagonal. Diagonal tensors have half the number of indices, i.e., a mode-2 diagonal tensor has one index.
  \item[scheme] The coupling scheme of the tensor. There are multiple ways to couple the angular momenta of the tensor indices. By default, the angular momenta of the first two creator indices are coupled, the resulting angular momentum is coupled with the third, etc., until all angular momenta have been coupled, and the process is repeated for the annihilator indices. This key accepts nested tuples that specify the coupling order of the tensor indices. Creator indices are numbered from 1 to $x$, annihilator indices from $x+1$ to $x+y$. The elements of each tuple are either tuples themselves or index numbers.
  Index numbers may be negated to request coupling of the time-reversed state.
  \item[latex] The LaTeX command used to typeset the tensor in the program output. By default, the name of the tensor is used.
\end{description}
To give an example,
\begin{amclisting}
declare X {
  mode = (2,2),
  scheme = ((1,-4),(3,-2)),
  scalar = true
}
\end{amclisting}
declares a scalar tensor $X$ with two creator and two annihilator indices, whose $m$-scheme, i.e. $SU(2)$ uncoupled, matrix elements can be recovered via
\begin{align}
  X_{pqrs} &= (-1)^{j_s-m_s+j_q-m_q}
\notag\\&\phantom{{}={}}\times
\sum_{JM} \clebsch{j_p}{m_p}{j_s}{-m_s}{J}{M}\clebsch{j_r}{m_r}{j_q}{-m_q}{J}{M}
  X^{J}_{\tilde p \tilde q \tilde r \tilde s}.
\end{align}
Equations are declared as
\begin{amclisting}
$variable$ = $expression$;
\end{amclisting}
The variable on the left-hand side is a declared tensor with index subscripts, such as \lstinline!X_pqrs!.
Indices consisting of more than one character can be used by enclosing the subscript with braces and separating the indices with spaces, like in \lstinline!X_{k1 k2 k3 k4}!.
Index names can consist of letters, numbers, and underscore characters.
The expression on the right-hand side consists of sums of products of tensor variables, denoted by \lstinline!+! and \lstinline!*! operators.
Two special operators are available: \lstinline!sum! and \lstinline!P!.
The \lstinline!sum! operator lists the indices to be summed over.
It is used in the following way:
\begin{amclisting}
Z_abcd = sum_pq(X_abpq * Y_pqcd);
\end{amclisting}
The subscript lists the indices.
The same rules apply as for tensor variables.
All indices have to be mentioned exactly once either on the left-hand side of the equation or in the subscript of the sum operator.

The other operator is the permutation operator \lstinline!P!.
It supports two modes of operation: used as $P(ij)$, it permutes indices $i$ and $j$ in the expression to its right.
Used as
\begin{align*}
P(i_1\dotsc i_m/j_1\dotsc j_n/\dotsc/k_1\dotsc k_p)\, ,
\end{align*}
it generates all distinct permutations between the index sets separated by slashes.
Concretely, $P(i/j) = 1 - P(ij)$, $P(ij/k) = 1 - P(ik) - P(jk)$, and $P(i/j/k) = 1 - P(ij) - P(ik) - P(jk) + P(ij)P(jk) + P(ik)P(jk)$.

As an example, an equation arising from the three-body part C3 of the commutator of a normal-ordered two-body operator A2 with a three-body operator B3, needed for the IMSRG(3), can be entered like this:
\begin{amclisting}
C3_pqrstu = 1/2 * sum_ab(
  (nbar_a*nbar_b - n_a*n_b) *
  (P(pq/r)*A2_pqab*B3_abrstu
    - P(st/u)*B3_pqrabu*A2_abst)
  + (nbar_a*n_b - n_a*nbar_b) *
    P(pq/r) * P(st/u) *
    B3_pqastb * A2_brau);
\end{amclisting}
The tensors n and nbar are diagonal one-body tensors containing occupation numbers.

\section{Organization of the code}

The \AMC{} code is organized into five modules: \texttt{ast}, \texttt{output}, \texttt{parser}, \texttt{reduction}, and \texttt{yutsis}.
The \texttt{ast} module defines classes whose instances make up the abstract syntax trees that are processed by the package, along with some helper classes that simplify working with the trees themselves.
The \texttt{output} module contains functions that turn abstract syntax trees back into other formats.
Currently, it only contains a module for LaTeX output.
The \texttt{parser} module provides a parser based on the PLY parser generator \cite{PLY} that produces abstract syntax trees from \AMC{} files.
The \texttt{reduction} module contains functions to perform the angular-mo\-men\-tum reduction itself.
Finally, the \texttt{yutsis} module contains classes and functions for building and manipulating Yutsis graphs, as well as for simplifying the resulting expressions.

The \AMC{} package is directly executable, and the installer creates a wrapper named \amc{} for convenience.
Executing the package provides a command-line interface for parsing an \AMC{} file, reduction of the contained expressions, and output of a LaTeX file.

The program flow is the following:
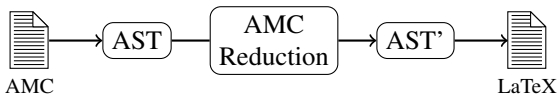
\begin{center}
\includegrtex{flow}
\end{center}
First, the \AMC{} file is parsed into an abstract syntax tree. The tree of each equation is expanded until it consists of a sum of products. For each term, a Yutsis graph is constructed according to the coupling schemes of the mentioned tensors. The reduction procedure looks for 2-, 3- and 4-cycles in the graph and applies the rules discussed in Sec.~\ref{sec:factorization}, iteratively factorizing the graph until it is completely expressed in terms of Kronecker deltas, triangular deltas, and $6j$-symbols. If enabled, a post-processing step tries to reconstruct $9j$-symbols by combining sets of three $6j$-symbols. The resulting abstract syntax tree is constructed by replacing all tensor variables with reduced ones and adding the objects resulting from the reduction of the Yutsis graph. This syntax tree represents the reduced equation, and is subsequently converted to a LaTeX expression and written to the output document.

\subsection{Testing files}

The \AMC{} package contains 7 example input files along with the outputs generated by the \amc{} program. The examples cover all applications discussed in \autoref{sec:appli}. Additionally, a more complex example is provided in the form of commutators of three-body operators that appear in IMSRG(3), and a file showing how to derive the Pandya transform of a scalar and a non-scalar tensor in a few lines of \AMC{} code.

%
%

\subsection{Methods}
In this section only a pointer to the central methods is provided.
See the API documentation that accompanies the package for more information.

\prog{parser.Parser.parse}{(instance method)
Parse a string into an abstract syntax tree according to the \AMC{} language grammar.
}
\prog{reduction.reduce\_equation}{
Reduce an equation, given as an abstract syntax tree, to symmetry-restricted form.
}
\prog{output.latex.equations\_to\_document}{
Turn a list of equations into a LaTeX document.
The equations can be in symmetry-reduced or unreduced form.
}

\begin{acknowledgements}
We thank Jan Hoppe and Matthias Heinz for proofreading the document and troubleshooting the installation.
This publication is based on work supported in part by the framework of the Espace de Structure et de r\'eactions Nucl\'eaires Th\'eorique (ESNT) at CEA, and the  Deutsche  Forschungsgemeinschaft  (DFG,  German Research Foundation) – Projektnummer 279384907 – SFB 1245.
This work has been supported by the U.S. Department of Energy,
Office of Science, Office of Nuclear Physics under award DE-SC0017887.

\end{acknowledgements}

\appendix
\section{Fundamentals of graph theory}
As alluded before the core concept of angular-momentum network is the correspondence between $3j$ symbols and vertices with three attached lines that are joined with each other by contracting common angular-momentum states.
The proper mathematical description is given in terms of graphs.

A \emph{graph} is a triplet $G=(E,V,I)$ consisting of a set of \emph{vertices} $V$ and a set of \emph{edges} $E$ with an \emph{incidence relation} $I$ specifying which vertices are connected via which edges. In this paper the sets $V$ and $E$ are assumed to be finite.
Let $v_1,v_2 \in V$ be two distinct vertices, then $v_1$ and $v_2$ are called \emph{adjacent} if there is a edge $e\in E$ connecting $v_1$ and $v_2$.
Additionally, an edge $e\in E$ is called \emph{incident} to $v$ if it starts or ends at $v$.
Given a vertex $v \in V$ its \emph{degree} $\text{deg}(v)$ denotes the number of incident edges.
If all vertices $v\in V$ $\text{deg}(v)=k$ then the graph is called \emph{$k$-regular}.
In the special case of $3$-regularity the graph is called \emph{cubic}.

Starting from a general string of of coupling symbols every $3j$-symbol yields a vertex of degree three in the graph.
Performing all contractions, i.e., joining disjoint vertices that have a common angular-momentum quantum number one obtains a connected graph.
Since every column of a $3j$ symbol corresponds to an incident edge the final graph is cubic.

\bibliographystyle{apsrev4-2}

\begingroup
\newcommand{\urlprefix}{}
\bibliography{short,clean}

\begin{thebibliography}{55}%
\makeatletter
\providecommand \@ifxundefined [1]{%
 \@ifx{#1\undefined}
}%
\providecommand \@ifnum [1]{%
 \ifnum #1\expandafter \@firstoftwo
 \else \expandafter \@secondoftwo
 \fi
}%
\providecommand \@ifx [1]{%
 \ifx #1\expandafter \@firstoftwo
 \else \expandafter \@secondoftwo
 \fi
}%
\providecommand \natexlab [1]{#1}%
\providecommand \enquote  [1]{``#1''}%
\providecommand \bibnamefont  [1]{#1}%
\providecommand \bibfnamefont [1]{#1}%
\providecommand \citenamefont [1]{#1}%
\providecommand \href@noop [0]{\@secondoftwo}%
\providecommand \href [0]{\begingroup \@sanitize@url \@href}%
\providecommand \@href[1]{\@@startlink{#1}\@@href}%
\providecommand \@@href[1]{\endgroup#1\@@endlink}%
\providecommand \@sanitize@url [0]{\catcode `\\12\catcode `\$12\catcode
  `\&12\catcode `\#12\catcode `\^12\catcode `\_12\catcode `\%12\relax}%
\providecommand \@@startlink[1]{}%
\providecommand \@@endlink[0]{}%
\providecommand \url  [0]{\begingroup\@sanitize@url \@url }%
\providecommand \@url [1]{\endgroup\@href {#1}{\urlprefix }}%
\providecommand \urlprefix  [0]{URL }%
\providecommand \Eprint [0]{\href }%
\providecommand \doibase [0]{https://doi.org/}%
\providecommand \selectlanguage [0]{\@gobble}%
\providecommand \bibinfo  [0]{\@secondoftwo}%
\providecommand \bibfield  [0]{\@secondoftwo}%
\providecommand \translation [1]{[#1]}%
\providecommand \BibitemOpen [0]{}%
\providecommand \bibitemStop [0]{}%
\providecommand \bibitemNoStop [0]{.\EOS\space}%
\providecommand \EOS [0]{\spacefactor3000\relax}%
\providecommand \BibitemShut  [1]{\csname bibitem#1\endcsname}%
\let\auto@bib@innerbib\@empty
\bibitem [{\citenamefont {Gezerlis}\ \emph {et~al.}(2013)\citenamefont
  {Gezerlis}, \citenamefont {Tews}, \citenamefont {Epelbaum}, \citenamefont
  {Gandolfi}, \citenamefont {Hebeler}, \citenamefont {Nogga},\ and\
  \citenamefont {Schwenk}}]{Gezerlis2013}%
  \BibitemOpen
  \bibfield  {author} {\bibinfo {author} {\bibfnamefont {A.}~\bibnamefont
  {Gezerlis}}, \bibinfo {author} {\bibfnamefont {I.}~\bibnamefont {Tews}},
  \bibinfo {author} {\bibfnamefont {E.}~\bibnamefont {Epelbaum}}, \bibinfo
  {author} {\bibfnamefont {S.}~\bibnamefont {Gandolfi}}, \bibinfo {author}
  {\bibfnamefont {K.}~\bibnamefont {Hebeler}}, \bibinfo {author} {\bibfnamefont
  {A.}~\bibnamefont {Nogga}},\ and\ \bibinfo {author} {\bibfnamefont
  {A.}~\bibnamefont {Schwenk}},\ }\href
  {https://doi.org/10.1103/PhysRevLett.111.032501} {\bibfield  {journal}
  {\bibinfo  {journal} {Phys. Rev. Lett.}\ }\textbf {\bibinfo {volume} {111}},\
  \bibinfo {pages} {032501} (\bibinfo {year} {2013})}\BibitemShut {NoStop}%
\bibitem [{\citenamefont {Carlson}\ \emph {et~al.}(2015)\citenamefont
  {Carlson}, \citenamefont {Gandolfi}, \citenamefont {Pederiva}, \citenamefont
  {Pieper}, \citenamefont {Schiavilla}, \citenamefont {Schmidt},\ and\
  \citenamefont {Wiringa}}]{Carlson:2015}%
  \BibitemOpen
  \bibfield  {author} {\bibinfo {author} {\bibfnamefont {J.}~\bibnamefont
  {Carlson}}, \bibinfo {author} {\bibfnamefont {S.}~\bibnamefont {Gandolfi}},
  \bibinfo {author} {\bibfnamefont {F.}~\bibnamefont {Pederiva}}, \bibinfo
  {author} {\bibfnamefont {S.~C.}\ \bibnamefont {Pieper}}, \bibinfo {author}
  {\bibfnamefont {R.}~\bibnamefont {Schiavilla}}, \bibinfo {author}
  {\bibfnamefont {K.~E.}\ \bibnamefont {Schmidt}},\ and\ \bibinfo {author}
  {\bibfnamefont {R.~B.}\ \bibnamefont {Wiringa}},\ }\href
  {https://doi.org/10.1103/RevModPhys.87.1067} {\bibfield  {journal} {\bibinfo
  {journal} {Rev. Mod. Phys.}\ }\textbf {\bibinfo {volume} {87}},\ \bibinfo
  {pages} {1067} (\bibinfo {year} {2015})}\BibitemShut {NoStop}%
\bibitem [{\citenamefont {Lynn}\ \emph {et~al.}(2017)\citenamefont {Lynn},
  \citenamefont {Tews}, \citenamefont {Carlson}, \citenamefont {Gandolfi},
  \citenamefont {Gezerlis}, \citenamefont {Schmidt},\ and\ \citenamefont
  {Schwenk}}]{Lynn2017}%
  \BibitemOpen
  \bibfield  {author} {\bibinfo {author} {\bibfnamefont {J.~E.}\ \bibnamefont
  {Lynn}}, \bibinfo {author} {\bibfnamefont {I.}~\bibnamefont {Tews}}, \bibinfo
  {author} {\bibfnamefont {J.}~\bibnamefont {Carlson}}, \bibinfo {author}
  {\bibfnamefont {S.}~\bibnamefont {Gandolfi}}, \bibinfo {author}
  {\bibfnamefont {A.}~\bibnamefont {Gezerlis}}, \bibinfo {author}
  {\bibfnamefont {K.~E.}\ \bibnamefont {Schmidt}},\ and\ \bibinfo {author}
  {\bibfnamefont {A.}~\bibnamefont {Schwenk}},\ }\href
  {https://doi.org/10.1103/PhysRevC.96.054007} {\bibfield  {journal} {\bibinfo
  {journal} {Phys. Rev. C}\ }\textbf {\bibinfo {volume} {96}},\ \bibinfo
  {pages} {054007} (\bibinfo {year} {2017})}\BibitemShut {NoStop}%
\bibitem [{\citenamefont {Lynn}\ \emph {et~al.}(2019)\citenamefont {Lynn},
  \citenamefont {Tews}, \citenamefont {Gandolfi},\ and\ \citenamefont
  {Lovato}}]{Lynn2019}%
  \BibitemOpen
  \bibfield  {author} {\bibinfo {author} {\bibfnamefont {J.}~\bibnamefont
  {Lynn}}, \bibinfo {author} {\bibfnamefont {I.}~\bibnamefont {Tews}}, \bibinfo
  {author} {\bibfnamefont {S.}~\bibnamefont {Gandolfi}},\ and\ \bibinfo
  {author} {\bibfnamefont {A.}~\bibnamefont {Lovato}},\ }\href
  {https://doi.org/10.1146/annurev-nucl-101918-023600} {\bibfield  {journal}
  {\bibinfo  {journal} {Annu. Rev. Nucl. Part. Sci.}\ }\textbf {\bibinfo
  {volume} {69}},\ \bibinfo {pages} {279} (\bibinfo {year} {2019})}\BibitemShut
  {NoStop}%
\bibitem [{\citenamefont {Navr{\'{a}}til}\ \emph {et~al.}(2009)\citenamefont
  {Navr{\'{a}}til}, \citenamefont {Quaglioni}, \citenamefont {Stetcu},\ and\
  \citenamefont {Barrett}}]{NaQu09}%
  \BibitemOpen
  \bibfield  {author} {\bibinfo {author} {\bibfnamefont {P.}~\bibnamefont
  {Navr{\'{a}}til}}, \bibinfo {author} {\bibfnamefont {S.}~\bibnamefont
  {Quaglioni}}, \bibinfo {author} {\bibfnamefont {I.}~\bibnamefont {Stetcu}},\
  and\ \bibinfo {author} {\bibfnamefont {B.}~\bibnamefont {Barrett}},\ }\href
  {https://doi.org/10.1088/0954-3899/36/8/083101} {\bibfield  {journal}
  {\bibinfo  {journal} {J. Phys. G}\ }\textbf {\bibinfo {volume} {36}},\
  \bibinfo {pages} {83101} (\bibinfo {year} {2009})}\BibitemShut {NoStop}%
\bibitem [{\citenamefont {Barrett}\ \emph {et~al.}(2013)\citenamefont
  {Barrett}, \citenamefont {Navr{\'{a}}til},\ and\ \citenamefont
  {Vary}}]{BaVa13}%
  \BibitemOpen
  \bibfield  {author} {\bibinfo {author} {\bibfnamefont {B.~R.}\ \bibnamefont
  {Barrett}}, \bibinfo {author} {\bibfnamefont {P.}~\bibnamefont
  {Navr{\'{a}}til}},\ and\ \bibinfo {author} {\bibfnamefont {J.~P.}\
  \bibnamefont {Vary}},\ }\href {https://doi.org/10.1016/j.ppnp.2012.10.003}
  {\bibfield  {journal} {\bibinfo  {journal} {Prog. Part. Nucl. Phys.}\
  }\textbf {\bibinfo {volume} {69}},\ \bibinfo {pages} {131} (\bibinfo {year}
  {2013})}\BibitemShut {NoStop}%
\bibitem [{\citenamefont {Goldstone}(1957)}]{Go57}%
  \BibitemOpen
  \bibfield  {author} {\bibinfo {author} {\bibfnamefont {J.}~\bibnamefont
  {Goldstone}},\ }\href {https://doi.org/10.1098/rspa.1957.0037} {\bibfield
  {journal} {\bibinfo  {journal} {Proc. Roy. Soc. A}\ }\textbf {\bibinfo
  {volume} {239}},\ \bibinfo {pages} {267} (\bibinfo {year}
  {1957})}\BibitemShut {NoStop}%
\bibitem [{\citenamefont {Shavitt}\ and\ \citenamefont
  {Bartlett}(2009)}]{Shavitt2009}%
  \BibitemOpen
  \bibfield  {author} {\bibinfo {author} {\bibfnamefont {I.}~\bibnamefont
  {Shavitt}}\ and\ \bibinfo {author} {\bibfnamefont {R.~J.}\ \bibnamefont
  {Bartlett}},\ }\href
  {http://www.amazon.com/Many-Body-Methods-Chemistry-Physics-Coupled-Cluster/dp/052181832X}
  {\emph {\bibinfo {title} {{Many-Body Methods in Chemistry and Physics: MBPT
  and Coupled-Cluster Theory (Cambridge Molecular Science)}}}}\ (\bibinfo
  {publisher} {Cambridge University Press},\ \bibinfo {year}
  {2009})\BibitemShut {NoStop}%
\bibitem [{\citenamefont {Tichai}\ \emph {et~al.}(2016)\citenamefont {Tichai},
  \citenamefont {Langhammer}, \citenamefont {Binder},\ and\ \citenamefont
  {Roth}}]{Tichai2016}%
  \BibitemOpen
  \bibfield  {author} {\bibinfo {author} {\bibfnamefont {A.}~\bibnamefont
  {Tichai}}, \bibinfo {author} {\bibfnamefont {J.}~\bibnamefont {Langhammer}},
  \bibinfo {author} {\bibfnamefont {S.}~\bibnamefont {Binder}},\ and\ \bibinfo
  {author} {\bibfnamefont {R.}~\bibnamefont {Roth}},\ }\href
  {https://doi.org/10.1016/j.physletb.2016.03.029} {\bibfield  {journal}
  {\bibinfo  {journal} {Phys. Lett. B}\ }\textbf {\bibinfo {volume} {756}},\
  \bibinfo {pages} {283} (\bibinfo {year} {2016})}\BibitemShut {NoStop}%
\bibitem [{\citenamefont {Hu}\ \emph {et~al.}(2016)\citenamefont {Hu},
  \citenamefont {Xu}, \citenamefont {Sun}, \citenamefont {Vary},\ and\
  \citenamefont {Li}}]{Hu:2016txm}%
  \BibitemOpen
  \bibfield  {author} {\bibinfo {author} {\bibfnamefont {B.~S.}\ \bibnamefont
  {Hu}}, \bibinfo {author} {\bibfnamefont {F.~R.}\ \bibnamefont {Xu}}, \bibinfo
  {author} {\bibfnamefont {Z.~H.}\ \bibnamefont {Sun}}, \bibinfo {author}
  {\bibfnamefont {J.~P.}\ \bibnamefont {Vary}},\ and\ \bibinfo {author}
  {\bibfnamefont {T.}~\bibnamefont {Li}},\ }\href
  {https://doi.org/10.1103/PhysRevC.94.014303} {\bibfield  {journal} {\bibinfo
  {journal} {Phys. Rev. C}\ }\textbf {\bibinfo {volume} {94}},\ \bibinfo
  {pages} {014303} (\bibinfo {year} {2016})}\BibitemShut {NoStop}%
\bibitem [{\citenamefont {Xu}\ \emph {et~al.}(2017)\citenamefont {Xu},
  \citenamefont {Hu},\ and\ \citenamefont {Sun}}]{Xu17}%
  \BibitemOpen
  \bibfield  {author} {\bibinfo {author} {\bibfnamefont {F.~R.}\ \bibnamefont
  {Xu}}, \bibinfo {author} {\bibfnamefont {B.~S.}\ \bibnamefont {Hu}},\ and\
  \bibinfo {author} {\bibfnamefont {Z.~H.}\ \bibnamefont {Sun}},\ }\bibinfo
  {title} {{Nuclear Many-Body Perturbation Calculations Based on the Chiral
  N\textsuperscript3LO Potential}},\ in\ \href
  {https://doi.org/10.1142/9789813229426_0081} {\emph {\bibinfo {booktitle}
  {Fission and Properties of Neutron-Rich Nuclei}}}\ (\bibinfo  {publisher}
  {World Scientific Publishing Company},\ \bibinfo {year} {2017})\ pp.\
  \bibinfo {pages} {438--445}\BibitemShut {NoStop}%
\bibitem [{\citenamefont {Tichai}\ \emph
  {et~al.}(2018{\natexlab{a}})\citenamefont {Tichai}, \citenamefont {Arthuis},
  \citenamefont {Duguet}, \citenamefont {Hergert}, \citenamefont {Som{\'{a}}},\
  and\ \citenamefont {Roth}}]{Tichai:2018mll}%
  \BibitemOpen
  \bibfield  {author} {\bibinfo {author} {\bibfnamefont {A.}~\bibnamefont
  {Tichai}}, \bibinfo {author} {\bibfnamefont {P.}~\bibnamefont {Arthuis}},
  \bibinfo {author} {\bibfnamefont {T.}~\bibnamefont {Duguet}}, \bibinfo
  {author} {\bibfnamefont {H.}~\bibnamefont {Hergert}}, \bibinfo {author}
  {\bibfnamefont {V.}~\bibnamefont {Som{\'{a}}}},\ and\ \bibinfo {author}
  {\bibfnamefont {R.}~\bibnamefont {Roth}},\ }\href
  {https://doi.org/10.1016/j.physletb.2018.09.044} {\bibfield  {journal}
  {\bibinfo  {journal} {Phys. Lett. B}\ }\textbf {\bibinfo {volume} {786}},\
  \bibinfo {pages} {195} (\bibinfo {year} {2018}{\natexlab{a}})}\BibitemShut
  {NoStop}%
\bibitem [{\citenamefont {Tichai}\ \emph
  {et~al.}(2018{\natexlab{b}})\citenamefont {Tichai}, \citenamefont
  {Gebrerufael}, \citenamefont {Vobig},\ and\ \citenamefont
  {Roth}}]{Tichai:2018ncsmpt}%
  \BibitemOpen
  \bibfield  {author} {\bibinfo {author} {\bibfnamefont {A.}~\bibnamefont
  {Tichai}}, \bibinfo {author} {\bibfnamefont {E.}~\bibnamefont {Gebrerufael}},
  \bibinfo {author} {\bibfnamefont {K.}~\bibnamefont {Vobig}},\ and\ \bibinfo
  {author} {\bibfnamefont {R.}~\bibnamefont {Roth}},\ }\href
  {https://doi.org/10.1016/j.physletb.2018.10.029} {\bibfield  {journal}
  {\bibinfo  {journal} {Phys. Lett. B}\ }\textbf {\bibinfo {volume} {786}},\
  \bibinfo {pages} {448} (\bibinfo {year} {2018}{\natexlab{b}})}\BibitemShut
  {NoStop}%
\bibitem [{\citenamefont {Hu}\ \emph {et~al.}(2018)\citenamefont {Hu},
  \citenamefont {Li},\ and\ \citenamefont {Xu}}]{Hu18arxiv}%
  \BibitemOpen
  \bibfield  {author} {\bibinfo {author} {\bibfnamefont {B.~S.}\ \bibnamefont
  {Hu}}, \bibinfo {author} {\bibfnamefont {T.}~\bibnamefont {Li}},\ and\
  \bibinfo {author} {\bibfnamefont {F.~R.}\ \bibnamefont {Xu}},\ }\Eprint
  {https://arxiv.org/abs/1810.08804} {arXiv:1810.08804 [nucl-th]}  (\bibinfo
  {year} {2018})\BibitemShut {NoStop}%
\bibitem [{\citenamefont {Tichai}\ \emph {et~al.}(2020)\citenamefont {Tichai},
  \citenamefont {Roth},\ and\ \citenamefont {Duguet}}]{Tichai:2020dna}%
  \BibitemOpen
  \bibfield  {author} {\bibinfo {author} {\bibfnamefont {A.}~\bibnamefont
  {Tichai}}, \bibinfo {author} {\bibfnamefont {R.}~\bibnamefont {Roth}},\ and\
  \bibinfo {author} {\bibfnamefont {T.}~\bibnamefont {Duguet}}\ }(\bibinfo
  {year} {2020})\ \Eprint {https://arxiv.org/abs/2001.10433} {arXiv:2001.10433
  [nucl-th]} \BibitemShut {NoStop}%
\bibitem [{\citenamefont {Demol}\ \emph {et~al.}(2020)\citenamefont {Demol},
  \citenamefont {Frosini}, \citenamefont {Tichai}, \citenamefont {Som\`{a}},\
  and\ \citenamefont {Duguet}}]{Demol2020a}%
  \BibitemOpen
  \bibfield  {author} {\bibinfo {author} {\bibfnamefont {P.}~\bibnamefont
  {Demol}}, \bibinfo {author} {\bibfnamefont {M.}~\bibnamefont {Frosini}},
  \bibinfo {author} {\bibfnamefont {A.}~\bibnamefont {Tichai}}, \bibinfo
  {author} {\bibfnamefont {V.}~\bibnamefont {Som\`{a}}},\ and\ \bibinfo
  {author} {\bibfnamefont {T.}~\bibnamefont {Duguet}}\ }(\bibinfo {year}
  {2020})\ \Eprint {https://arxiv.org/abs/2002.02724} {arXiv:2002.02724
  [nucl-th]} \BibitemShut {NoStop}%
\bibitem [{\citenamefont {Kowalski}\ \emph {et~al.}(2004)\citenamefont
  {Kowalski}, \citenamefont {Dean}, \citenamefont {Hjorth-Jensen},
  \citenamefont {Papenbrock},\ and\ \citenamefont {Piecuch}}]{KoDe04}%
  \BibitemOpen
  \bibfield  {author} {\bibinfo {author} {\bibfnamefont {K.}~\bibnamefont
  {Kowalski}}, \bibinfo {author} {\bibfnamefont {D.~J.}\ \bibnamefont {Dean}},
  \bibinfo {author} {\bibfnamefont {M.}~\bibnamefont {Hjorth-Jensen}}, \bibinfo
  {author} {\bibfnamefont {T.}~\bibnamefont {Papenbrock}},\ and\ \bibinfo
  {author} {\bibfnamefont {P.}~\bibnamefont {Piecuch}},\ }\href
  {https://doi.org/10.1103/PhysRevLett.92.132501} {\bibfield  {journal}
  {\bibinfo  {journal} {Phys. Rev. Lett.}\ }\textbf {\bibinfo {volume} {92}},\
  \bibinfo {pages} {132501} (\bibinfo {year} {2004})}\BibitemShut {NoStop}%
\bibitem [{\citenamefont {Binder}\ \emph {et~al.}(2013)\citenamefont {Binder},
  \citenamefont {Piecuch}, \citenamefont {Calci}, \citenamefont {Langhammer},
  \citenamefont {Navr{\'{a}}til},\ and\ \citenamefont {Roth}}]{Binder2013}%
  \BibitemOpen
  \bibfield  {author} {\bibinfo {author} {\bibfnamefont {S.}~\bibnamefont
  {Binder}}, \bibinfo {author} {\bibfnamefont {P.}~\bibnamefont {Piecuch}},
  \bibinfo {author} {\bibfnamefont {A.}~\bibnamefont {Calci}}, \bibinfo
  {author} {\bibfnamefont {J.}~\bibnamefont {Langhammer}}, \bibinfo {author}
  {\bibfnamefont {P.}~\bibnamefont {Navr{\'{a}}til}},\ and\ \bibinfo {author}
  {\bibfnamefont {R.}~\bibnamefont {Roth}},\ }\href
  {https://doi.org/10.1103/PhysRevC.88.054319} {\bibfield  {journal} {\bibinfo
  {journal} {Phys. Rev. C}\ }\textbf {\bibinfo {volume} {88}},\ \bibinfo
  {pages} {054319} (\bibinfo {year} {2013})}\BibitemShut {NoStop}%
\bibitem [{\citenamefont {Jansen}\ \emph {et~al.}(2014)\citenamefont {Jansen},
  \citenamefont {Engel}, \citenamefont {Hagen}, \citenamefont
  {Navr{\'{a}}til},\ and\ \citenamefont {Signoracci}}]{Ja14}%
  \BibitemOpen
  \bibfield  {author} {\bibinfo {author} {\bibfnamefont {G.~R.}\ \bibnamefont
  {Jansen}}, \bibinfo {author} {\bibfnamefont {J.}~\bibnamefont {Engel}},
  \bibinfo {author} {\bibfnamefont {G.}~\bibnamefont {Hagen}}, \bibinfo
  {author} {\bibfnamefont {P.}~\bibnamefont {Navr{\'{a}}til}},\ and\ \bibinfo
  {author} {\bibfnamefont {A.}~\bibnamefont {Signoracci}},\ }\href
  {https://doi.org/10.1103/PhysRevLett.113.142502} {\bibfield  {journal}
  {\bibinfo  {journal} {Phys. Rev. Lett.}\ }\textbf {\bibinfo {volume} {113}},\
  \bibinfo {pages} {142502} (\bibinfo {year} {2014})}\BibitemShut {NoStop}%
\bibitem [{\citenamefont {Signoracci}\ \emph {et~al.}(2015)\citenamefont
  {Signoracci}, \citenamefont {Duguet}, \citenamefont {Hagen},\ and\
  \citenamefont {Jansen}}]{Si15}%
  \BibitemOpen
  \bibfield  {author} {\bibinfo {author} {\bibfnamefont {A.}~\bibnamefont
  {Signoracci}}, \bibinfo {author} {\bibfnamefont {T.}~\bibnamefont {Duguet}},
  \bibinfo {author} {\bibfnamefont {G.}~\bibnamefont {Hagen}},\ and\ \bibinfo
  {author} {\bibfnamefont {G.~R.}\ \bibnamefont {Jansen}},\ }\href
  {https://doi.org/10.1103/PhysRevC.91.064320} {\bibfield  {journal} {\bibinfo
  {journal} {Phys. Rev. C}\ }\textbf {\bibinfo {volume} {91}},\ \bibinfo
  {pages} {064320} (\bibinfo {year} {2015})}\BibitemShut {NoStop}%
\bibitem [{\citenamefont {Duguet}(2015)}]{Duguet:2014jja}%
  \BibitemOpen
  \bibfield  {author} {\bibinfo {author} {\bibfnamefont {T.}~\bibnamefont
  {Duguet}},\ }\href {https://doi.org/10.1088/0954-3899/42/2/025107} {\bibfield
   {journal} {\bibinfo  {journal} {J. Phys. G}\ }\textbf {\bibinfo {volume}
  {42}},\ \bibinfo {pages} {025107} (\bibinfo {year} {2015})}\BibitemShut
  {NoStop}%
\bibitem [{\citenamefont {Duguet}\ and\ \citenamefont
  {Signoracci}(2017)}]{Duguet:2015yle}%
  \BibitemOpen
  \bibfield  {author} {\bibinfo {author} {\bibfnamefont {T.}~\bibnamefont
  {Duguet}}\ and\ \bibinfo {author} {\bibfnamefont {A.}~\bibnamefont
  {Signoracci}},\ }\href {https://doi.org/10.1088/0954-3899/44/1/015103}
  {\bibfield  {journal} {\bibinfo  {journal} {J. Phys. G}\ }\textbf {\bibinfo
  {volume} {44}},\ \bibinfo {pages} {015103} (\bibinfo {year}
  {2017})}\BibitemShut {NoStop}%
\bibitem [{\citenamefont {Qiu}\ \emph {et~al.}(2019)\citenamefont {Qiu},
  \citenamefont {Henderson}, \citenamefont {Duguet},\ and\ \citenamefont
  {Scuseria}}]{Qiu:2019edx}%
  \BibitemOpen
  \bibfield  {author} {\bibinfo {author} {\bibfnamefont {Y.}~\bibnamefont
  {Qiu}}, \bibinfo {author} {\bibfnamefont {T.~M.}\ \bibnamefont {Henderson}},
  \bibinfo {author} {\bibfnamefont {T.}~\bibnamefont {Duguet}},\ and\ \bibinfo
  {author} {\bibfnamefont {G.~E.}\ \bibnamefont {Scuseria}},\ }\href
  {https://doi.org/10.1103/PhysRevC.99.044301} {\bibfield  {journal} {\bibinfo
  {journal} {Phys. Rev. C}\ }\textbf {\bibinfo {volume} {99}},\ \bibinfo
  {pages} {044301} (\bibinfo {year} {2019})}\BibitemShut {NoStop}%
\bibitem [{\citenamefont {Dickhoff}\ and\ \citenamefont
  {Barbieri}(2004)}]{Dickhoff:2004xx}%
  \BibitemOpen
  \bibfield  {author} {\bibinfo {author} {\bibfnamefont {W.~H.}\ \bibnamefont
  {Dickhoff}}\ and\ \bibinfo {author} {\bibfnamefont {C.}~\bibnamefont
  {Barbieri}},\ }\href {https://doi.org/10.1016/j.ppnp.2004.02.038} {\bibfield
  {journal} {\bibinfo  {journal} {Prog. Part. Nucl. Phys.}\ }\textbf {\bibinfo
  {volume} {52}},\ \bibinfo {pages} {377} (\bibinfo {year} {2004})}\BibitemShut
  {NoStop}%
\bibitem [{\citenamefont {Cipollone}\ \emph {et~al.}(2013)\citenamefont
  {Cipollone}, \citenamefont {Barbieri},\ and\ \citenamefont
  {Navr{\'{a}}til}}]{CiBa13}%
  \BibitemOpen
  \bibfield  {author} {\bibinfo {author} {\bibfnamefont {A.}~\bibnamefont
  {Cipollone}}, \bibinfo {author} {\bibfnamefont {C.}~\bibnamefont
  {Barbieri}},\ and\ \bibinfo {author} {\bibfnamefont {P.}~\bibnamefont
  {Navr{\'{a}}til}},\ }\href {https://doi.org/10.1103/PhysRevLett.111.062501}
  {\bibfield  {journal} {\bibinfo  {journal} {Phys. Rev. Lett.}\ }\textbf
  {\bibinfo {volume} {111}},\ \bibinfo {pages} {062501} (\bibinfo {year}
  {2013})}\BibitemShut {NoStop}%
\bibitem [{\citenamefont {Carbone}\ \emph {et~al.}(2013)\citenamefont
  {Carbone}, \citenamefont {Cipollone}, \citenamefont {Barbieri}, \citenamefont
  {Rios},\ and\ \citenamefont {Polls}}]{Carbone:2013eqa}%
  \BibitemOpen
  \bibfield  {author} {\bibinfo {author} {\bibfnamefont {A.}~\bibnamefont
  {Carbone}}, \bibinfo {author} {\bibfnamefont {A.}~\bibnamefont {Cipollone}},
  \bibinfo {author} {\bibfnamefont {C.}~\bibnamefont {Barbieri}}, \bibinfo
  {author} {\bibfnamefont {A.}~\bibnamefont {Rios}},\ and\ \bibinfo {author}
  {\bibfnamefont {A.}~\bibnamefont {Polls}},\ }\href
  {https://doi.org/10.1103/PhysRevC.88.054326} {\bibfield  {journal} {\bibinfo
  {journal} {Phys. Rev. C}\ }\textbf {\bibinfo {volume} {88}},\ \bibinfo
  {pages} {054326} (\bibinfo {year} {2013})}\BibitemShut {NoStop}%
\bibitem [{\citenamefont {Som{\`{a}}}\ \emph {et~al.}(2011)\citenamefont
  {Som{\`{a}}}, \citenamefont {Duguet},\ and\ \citenamefont
  {Barbieri}}]{Soma:2011aj}%
  \BibitemOpen
  \bibfield  {author} {\bibinfo {author} {\bibfnamefont {V.}~\bibnamefont
  {Som{\`{a}}}}, \bibinfo {author} {\bibfnamefont {T.}~\bibnamefont {Duguet}},\
  and\ \bibinfo {author} {\bibfnamefont {C.}~\bibnamefont {Barbieri}},\ }\href
  {https://doi.org/10.1103/PhysRevC.84.064317} {\bibfield  {journal} {\bibinfo
  {journal} {Phys. Rev. C}\ }\textbf {\bibinfo {volume} {84}},\ \bibinfo
  {pages} {064317} (\bibinfo {year} {2011})}\BibitemShut {NoStop}%
\bibitem [{\citenamefont {Som{\`{a}}}\ \emph {et~al.}(2014)\citenamefont
  {Som{\`{a}}}, \citenamefont {Cipollone}, \citenamefont {Barbieri},
  \citenamefont {Navr{\'{a}}til},\ and\ \citenamefont {Duguet}}]{SoCi13}%
  \BibitemOpen
  \bibfield  {author} {\bibinfo {author} {\bibfnamefont {V.}~\bibnamefont
  {Som{\`{a}}}}, \bibinfo {author} {\bibfnamefont {A.}~\bibnamefont
  {Cipollone}}, \bibinfo {author} {\bibfnamefont {C.}~\bibnamefont {Barbieri}},
  \bibinfo {author} {\bibfnamefont {P.}~\bibnamefont {Navr{\'{a}}til}},\ and\
  \bibinfo {author} {\bibfnamefont {T.}~\bibnamefont {Duguet}},\ }\href
  {https://doi.org/10.1103/PhysRevC.89.061301} {\bibfield  {journal} {\bibinfo
  {journal} {Phys. Rev. C}\ }\textbf {\bibinfo {volume} {89}},\ \bibinfo
  {pages} {061301(R)} (\bibinfo {year} {2014})}\BibitemShut {NoStop}%
\bibitem [{\citenamefont {Tsukiyama}\ \emph {et~al.}(2011)\citenamefont
  {Tsukiyama}, \citenamefont {Bogner},\ and\ \citenamefont
  {Schwenk}}]{Tsukiyama:2011}%
  \BibitemOpen
  \bibfield  {author} {\bibinfo {author} {\bibfnamefont {K.}~\bibnamefont
  {Tsukiyama}}, \bibinfo {author} {\bibfnamefont {S.~K.}\ \bibnamefont
  {Bogner}},\ and\ \bibinfo {author} {\bibfnamefont {A.}~\bibnamefont
  {Schwenk}},\ }\href {https://doi.org/10.1103/PhysRevLett.106.222502}
  {\bibfield  {journal} {\bibinfo  {journal} {Phys. Rev. Lett.}\ }\textbf
  {\bibinfo {volume} {106}},\ \bibinfo {pages} {222502} (\bibinfo {year}
  {2011})}\BibitemShut {NoStop}%
\bibitem [{\citenamefont {Tsukiyama}\ \emph {et~al.}(2012)\citenamefont
  {Tsukiyama}, \citenamefont {Bogner},\ and\ \citenamefont
  {Schwenk}}]{Tsukiyama:2012}%
  \BibitemOpen
  \bibfield  {author} {\bibinfo {author} {\bibfnamefont {K.}~\bibnamefont
  {Tsukiyama}}, \bibinfo {author} {\bibfnamefont {S.~K.}\ \bibnamefont
  {Bogner}},\ and\ \bibinfo {author} {\bibfnamefont {A.}~\bibnamefont
  {Schwenk}},\ }\href {https://link.aps.org/doi/10.1103/PhysRevC.85.061304}
  {\bibfield  {journal} {\bibinfo  {journal} {Phys. Rev. C}\ }\textbf {\bibinfo
  {volume} {85}},\ \bibinfo {pages} {061304} (\bibinfo {year}
  {2012})}\BibitemShut {NoStop}%
\bibitem [{\citenamefont {Hergert}\ \emph {et~al.}(2013)\citenamefont
  {Hergert}, \citenamefont {Bogner}, \citenamefont {Binder}, \citenamefont
  {Calci}, \citenamefont {Langhammer}, \citenamefont {Roth},\ and\
  \citenamefont {Schwenk}}]{Hergert2013}%
  \BibitemOpen
  \bibfield  {author} {\bibinfo {author} {\bibfnamefont {H.}~\bibnamefont
  {Hergert}}, \bibinfo {author} {\bibfnamefont {S.~K.}\ \bibnamefont {Bogner}},
  \bibinfo {author} {\bibfnamefont {S.}~\bibnamefont {Binder}}, \bibinfo
  {author} {\bibfnamefont {a.}~\bibnamefont {Calci}}, \bibinfo {author}
  {\bibfnamefont {J.}~\bibnamefont {Langhammer}}, \bibinfo {author}
  {\bibfnamefont {R.}~\bibnamefont {Roth}},\ and\ \bibinfo {author}
  {\bibfnamefont {a.}~\bibnamefont {Schwenk}},\ }\href
  {https://doi.org/10.1103/PhysRevC.87.034307} {\bibfield  {journal} {\bibinfo
  {journal} {Phys. Rev. C}\ }\textbf {\bibinfo {volume} {87}},\ \bibinfo
  {pages} {034307} (\bibinfo {year} {2013})}\BibitemShut {NoStop}%
\bibitem [{\citenamefont {Bogner}\ \emph {et~al.}(2014)\citenamefont {Bogner},
  \citenamefont {Hergert}, \citenamefont {Holt}, \citenamefont {Schwenk},
  \citenamefont {Binder}, \citenamefont {Calci}, \citenamefont {Langhammer},\
  and\ \citenamefont {Roth}}]{Bo14}%
  \BibitemOpen
  \bibfield  {author} {\bibinfo {author} {\bibfnamefont {S.~K.}\ \bibnamefont
  {Bogner}}, \bibinfo {author} {\bibfnamefont {H.}~\bibnamefont {Hergert}},
  \bibinfo {author} {\bibfnamefont {J.~D.}\ \bibnamefont {Holt}}, \bibinfo
  {author} {\bibfnamefont {A.}~\bibnamefont {Schwenk}}, \bibinfo {author}
  {\bibfnamefont {S.}~\bibnamefont {Binder}}, \bibinfo {author} {\bibfnamefont
  {A.}~\bibnamefont {Calci}}, \bibinfo {author} {\bibfnamefont
  {J.}~\bibnamefont {Langhammer}},\ and\ \bibinfo {author} {\bibfnamefont
  {R.}~\bibnamefont {Roth}},\ }\href
  {https://doi.org/10.1103/PhysRevLett.113.142501} {\bibfield  {journal}
  {\bibinfo  {journal} {Phys. Rev. Lett.}\ }\textbf {\bibinfo {volume} {113}},\
  \bibinfo {pages} {142501} (\bibinfo {year} {2014})}\BibitemShut {NoStop}%
\bibitem [{\citenamefont {Hergert}\ \emph
  {et~al.}(2016{\natexlab{a}})\citenamefont {Hergert}, \citenamefont {Bogner},
  \citenamefont {Morris}, \citenamefont {Schwenk},\ and\ \citenamefont
  {Tsukiyama}}]{H15}%
  \BibitemOpen
  \bibfield  {author} {\bibinfo {author} {\bibfnamefont {H.}~\bibnamefont
  {Hergert}}, \bibinfo {author} {\bibfnamefont {S.~K.}\ \bibnamefont {Bogner}},
  \bibinfo {author} {\bibfnamefont {T.~D.}\ \bibnamefont {Morris}}, \bibinfo
  {author} {\bibfnamefont {A.}~\bibnamefont {Schwenk}},\ and\ \bibinfo {author}
  {\bibfnamefont {K.}~\bibnamefont {Tsukiyama}},\ }\href
  {https://doi.org/10.1016/j.physrep.2015.12.007} {\bibfield  {journal}
  {\bibinfo  {journal} {Phys. Rep.}\ }\textbf {\bibinfo {volume} {621}},\
  \bibinfo {pages} {165} (\bibinfo {year} {2016}{\natexlab{a}})}\BibitemShut
  {NoStop}%
\bibitem [{\citenamefont {Morris}\ \emph {et~al.}(2018)\citenamefont {Morris},
  \citenamefont {Simonis}, \citenamefont {Stroberg}, \citenamefont {Stumpf},
  \citenamefont {Hagen}, \citenamefont {Holt}, \citenamefont {Jansen},
  \citenamefont {Papenbrock}, \citenamefont {Roth},\ and\ \citenamefont
  {Schwenk}}]{Morris:2017vxi}%
  \BibitemOpen
  \bibfield  {author} {\bibinfo {author} {\bibfnamefont {T.~D.}\ \bibnamefont
  {Morris}}, \bibinfo {author} {\bibfnamefont {J.}~\bibnamefont {Simonis}},
  \bibinfo {author} {\bibfnamefont {S.~R.}\ \bibnamefont {Stroberg}}, \bibinfo
  {author} {\bibfnamefont {C.}~\bibnamefont {Stumpf}}, \bibinfo {author}
  {\bibfnamefont {G.}~\bibnamefont {Hagen}}, \bibinfo {author} {\bibfnamefont
  {J.~D.}\ \bibnamefont {Holt}}, \bibinfo {author} {\bibfnamefont {G.~R.}\
  \bibnamefont {Jansen}}, \bibinfo {author} {\bibfnamefont {T.}~\bibnamefont
  {Papenbrock}}, \bibinfo {author} {\bibfnamefont {R.}~\bibnamefont {Roth}},\
  and\ \bibinfo {author} {\bibfnamefont {A.}~\bibnamefont {Schwenk}},\ }\href
  {https://doi.org/10.1103/PhysRevLett.120.152503} {\bibfield  {journal}
  {\bibinfo  {journal} {Phys. Rev. Lett.}\ }\textbf {\bibinfo {volume} {120}},\
  \bibinfo {pages} {152503} (\bibinfo {year} {2018})}\BibitemShut {NoStop}%
\bibitem [{\citenamefont {Stroberg}\ \emph {et~al.}(2017)\citenamefont
  {Stroberg}, \citenamefont {Calci}, \citenamefont {Hergert}, \citenamefont
  {Holt}, \citenamefont {Bogner}, \citenamefont {Roth},\ and\ \citenamefont
  {Schwenk}}]{Stroberg2017}%
  \BibitemOpen
  \bibfield  {author} {\bibinfo {author} {\bibfnamefont {S.~R.}\ \bibnamefont
  {Stroberg}}, \bibinfo {author} {\bibfnamefont {A.}~\bibnamefont {Calci}},
  \bibinfo {author} {\bibfnamefont {H.}~\bibnamefont {Hergert}}, \bibinfo
  {author} {\bibfnamefont {J.~D.}\ \bibnamefont {Holt}}, \bibinfo {author}
  {\bibfnamefont {S.~K.}\ \bibnamefont {Bogner}}, \bibinfo {author}
  {\bibfnamefont {R.}~\bibnamefont {Roth}},\ and\ \bibinfo {author}
  {\bibfnamefont {A.}~\bibnamefont {Schwenk}},\ }\href
  {https://doi.org/10.1103/PhysRevLett.118.032502} {\bibfield  {journal}
  {\bibinfo  {journal} {Phys. Rev. Lett.}\ }\textbf {\bibinfo {volume} {118}},\
  \bibinfo {pages} {032502} (\bibinfo {year} {2017})}\BibitemShut {NoStop}%
\bibitem [{\citenamefont {Parzuchowski}\ \emph
  {et~al.}(2017{\natexlab{a}})\citenamefont {Parzuchowski}, \citenamefont
  {Morris},\ and\ \citenamefont {Bogner}}]{Parzuchowski2017}%
  \BibitemOpen
  \bibfield  {author} {\bibinfo {author} {\bibfnamefont {N.~M.}\ \bibnamefont
  {Parzuchowski}}, \bibinfo {author} {\bibfnamefont {T.~D.}\ \bibnamefont
  {Morris}},\ and\ \bibinfo {author} {\bibfnamefont {S.~K.}\ \bibnamefont
  {Bogner}},\ }\href {https://doi.org/10.1103/PhysRevC.95.044304} {\bibfield
  {journal} {\bibinfo  {journal} {Phys. Rev. C}\ }\textbf {\bibinfo {volume}
  {95}},\ \bibinfo {pages} {044304} (\bibinfo {year}
  {2017}{\natexlab{a}})}\BibitemShut {NoStop}%
\bibitem [{\citenamefont {Hergert}\ \emph {et~al.}(2018)\citenamefont
  {Hergert}, \citenamefont {Yao}, \citenamefont {Morris}, \citenamefont
  {Parzuchowski}, \citenamefont {Bogner},\ and\ \citenamefont
  {Engel}}]{Hergert:2018wmx}%
  \BibitemOpen
  \bibfield  {author} {\bibinfo {author} {\bibfnamefont {H.}~\bibnamefont
  {Hergert}}, \bibinfo {author} {\bibfnamefont {J.}~\bibnamefont {Yao}},
  \bibinfo {author} {\bibfnamefont {T.~D.}\ \bibnamefont {Morris}}, \bibinfo
  {author} {\bibfnamefont {N.~M.}\ \bibnamefont {Parzuchowski}}, \bibinfo
  {author} {\bibfnamefont {S.~K.}\ \bibnamefont {Bogner}},\ and\ \bibinfo
  {author} {\bibfnamefont {J.}~\bibnamefont {Engel}},\ }in\ \href@noop {}
  {\emph {\bibinfo {booktitle} {19th International Conference on Recent
  Progress in Many-Body Theories (RPMBT19) Pohang, Korea, June 25-30, 2017}}}\
  (\bibinfo {year} {2018})\ \Eprint {https://arxiv.org/abs/1805.09221}
  {arXiv:1805.09221 [nucl-th]} \BibitemShut {NoStop}%
\bibitem [{\citenamefont {Paldus}\ and\ \citenamefont
  {Wong}(1973)}]{Paldus1973}%
  \BibitemOpen
  \bibfield  {author} {\bibinfo {author} {\bibfnamefont {J.}~\bibnamefont
  {Paldus}}\ and\ \bibinfo {author} {\bibfnamefont {H.}~\bibnamefont {Wong}},\
  }\href {https://doi.org/10.1016/0010-4655(73)90016-7} {\bibfield  {journal}
  {\bibinfo  {journal} {Comp. Phys. Comm.}\ }\textbf {\bibinfo {volume} {6}},\
  \bibinfo {pages} {1} (\bibinfo {year} {1973})}\BibitemShut {NoStop}%
\bibitem [{\citenamefont {Kaldor}(1976)}]{Ka76}%
  \BibitemOpen
  \bibfield  {author} {\bibinfo {author} {\bibfnamefont {U.}~\bibnamefont
  {Kaldor}},\ }\href {https://doi.org/10.1016/0021-9991(76)90092-9} {\bibfield
  {journal} {\bibinfo  {journal} {J. Comp. Phys.}\ }\textbf {\bibinfo {volume}
  {20}},\ \bibinfo {pages} {432} (\bibinfo {year} {1976})}\BibitemShut
  {NoStop}%
\bibitem [{\citenamefont {Cs{\'{e}}pes}\ and\ \citenamefont
  {Pipek}(1988)}]{Csepes1988}%
  \BibitemOpen
  \bibfield  {author} {\bibinfo {author} {\bibfnamefont {Z.}~\bibnamefont
  {Cs{\'{e}}pes}}\ and\ \bibinfo {author} {\bibfnamefont {J.}~\bibnamefont
  {Pipek}},\ }\href {https://doi.org/10.1016/0021-9991(88)90153-2} {\bibfield
  {journal} {\bibinfo  {journal} {J. Comp. Phys.}\ }\textbf {\bibinfo {volume}
  {77}},\ \bibinfo {pages} {1} (\bibinfo {year} {1988})}\BibitemShut {NoStop}%
\bibitem [{\citenamefont {Lyons}\ \emph {et~al.}(1994)\citenamefont {Lyons},
  \citenamefont {Moncrieff},\ and\ \citenamefont {Wilson}}]{Lyons:1994ew}%
  \BibitemOpen
  \bibfield  {author} {\bibinfo {author} {\bibfnamefont {J.}~\bibnamefont
  {Lyons}}, \bibinfo {author} {\bibfnamefont {D.}~\bibnamefont {Moncrieff}},\
  and\ \bibinfo {author} {\bibfnamefont {S.}~\bibnamefont {Wilson}},\ }\href
  {https://doi.org/10.1016/0010-4655(94)90205-4} {\bibfield  {journal}
  {\bibinfo  {journal} {Comp. Phys. Comm.}\ }\textbf {\bibinfo {volume} {84}},\
  \bibinfo {pages} {91} (\bibinfo {year} {1994})}\BibitemShut {NoStop}%
\bibitem [{\citenamefont {Xiao}\ \emph {et~al.}(2013)\citenamefont {Xiao},
  \citenamefont {Wang},\ and\ \citenamefont {Zhu}}]{Xiao2013}%
  \BibitemOpen
  \bibfield  {author} {\bibinfo {author} {\bibfnamefont {B.}~\bibnamefont
  {Xiao}}, \bibinfo {author} {\bibfnamefont {H.}~\bibnamefont {Wang}},\ and\
  \bibinfo {author} {\bibfnamefont {S.-h.}\ \bibnamefont {Zhu}},\ }\href
  {https://doi.org/10.1016/j.cpc.2013.03.015} {\bibfield  {journal} {\bibinfo
  {journal} {Comp. Phys. Comm.}\ }\textbf {\bibinfo {volume} {184}},\ \bibinfo
  {pages} {1966} (\bibinfo {year} {2013})}\BibitemShut {NoStop}%
\bibitem [{\citenamefont {Hirata}(2003)}]{Hi03}%
  \BibitemOpen
  \bibfield  {author} {\bibinfo {author} {\bibfnamefont {S.}~\bibnamefont
  {Hirata}},\ }\href {https://doi.org/10.1021/jp034596z} {\bibfield  {journal}
  {\bibinfo  {journal} {J. Phys. Chem. A}\ }\textbf {\bibinfo {volume} {107}},\
  \bibinfo {pages} {9887} (\bibinfo {year} {2003})}\BibitemShut {NoStop}%
\bibitem [{\citenamefont {Arthuis}\ \emph {et~al.}(2019)\citenamefont
  {Arthuis}, \citenamefont {Duguet}, \citenamefont {Tichai}, \citenamefont
  {Lasseri},\ and\ \citenamefont {Ebran}}]{Arthuis:2018yoo}%
  \BibitemOpen
  \bibfield  {author} {\bibinfo {author} {\bibfnamefont {P.}~\bibnamefont
  {Arthuis}}, \bibinfo {author} {\bibfnamefont {T.}~\bibnamefont {Duguet}},
  \bibinfo {author} {\bibfnamefont {A.}~\bibnamefont {Tichai}}, \bibinfo
  {author} {\bibfnamefont {R.-D.}\ \bibnamefont {Lasseri}},\ and\ \bibinfo
  {author} {\bibfnamefont {J.-P.}\ \bibnamefont {Ebran}},\ }\href
  {https://doi.org/10.1016/j.cpc.2018.11.023} {\bibfield  {journal} {\bibinfo
  {journal} {Comp. Phys. Comm.}\ }\textbf {\bibinfo {volume} {240}},\ \bibinfo
  {pages} {202} (\bibinfo {year} {2019})}\BibitemShut {NoStop}%
\bibitem [{\citenamefont {Wormer}\ and\ \citenamefont
  {Paldus}(2006)}]{Wormer2006}%
  \BibitemOpen
  \bibfield  {author} {\bibinfo {author} {\bibfnamefont {P.~E.}\ \bibnamefont
  {Wormer}}\ and\ \bibinfo {author} {\bibfnamefont {J.}~\bibnamefont
  {Paldus}},\ }\href {https://doi.org/10.1016/S0065-3276(06)51002-0} {\bibfield
   {journal} {\bibinfo  {journal} {Adv. Quant. Chem.}\ }\textbf {\bibinfo
  {volume} {51}},\ \bibinfo {pages} {59} (\bibinfo {year} {2006})}\BibitemShut
  {NoStop}%
\bibitem [{\citenamefont {Dytrych}\ \emph {et~al.}(2008)\citenamefont
  {Dytrych}, \citenamefont {Sviratcheva}, \citenamefont {Draayer},
  \citenamefont {Bahri},\ and\ \citenamefont {Vary}}]{Dytrych2008}%
  \BibitemOpen
  \bibfield  {author} {\bibinfo {author} {\bibfnamefont {T.}~\bibnamefont
  {Dytrych}}, \bibinfo {author} {\bibfnamefont {K.~D.}\ \bibnamefont
  {Sviratcheva}}, \bibinfo {author} {\bibfnamefont {J.~P.}\ \bibnamefont
  {Draayer}}, \bibinfo {author} {\bibfnamefont {C.}~\bibnamefont {Bahri}},\
  and\ \bibinfo {author} {\bibfnamefont {J.~P.}\ \bibnamefont {Vary}},\ }\href
  {https://doi.org/10.1088/0954-3899/35/12/123101} {\bibfield  {journal}
  {\bibinfo  {journal} {J. Phys. G}\ }\textbf {\bibinfo {volume} {35}},\
  \bibinfo {pages} {123101} (\bibinfo {year} {2008})}\BibitemShut {NoStop}%
\bibitem [{\citenamefont {Varshalovich}\ \emph {et~al.}(1988)\citenamefont
  {Varshalovich}, \citenamefont {Moskalev},\ and\ \citenamefont
  {Khersonskii}}]{VaMo88}%
  \BibitemOpen
  \bibfield  {author} {\bibinfo {author} {\bibfnamefont {D.~A.}\ \bibnamefont
  {Varshalovich}}, \bibinfo {author} {\bibfnamefont {A.~N.}\ \bibnamefont
  {Moskalev}},\ and\ \bibinfo {author} {\bibfnamefont {V.~K.}\ \bibnamefont
  {Khersonskii}},\ }\href {https://doi.org/10.1142/0270} {\emph {\bibinfo
  {title} {{Quantum Theory of Angular Momentum}}}}\ (\bibinfo  {publisher}
  {World Scientific Publishing Company},\ \bibinfo {year} {1988})\BibitemShut
  {NoStop}%
\bibitem [{\citenamefont {Suhonen}(2007)}]{Su07}%
  \BibitemOpen
  \bibfield  {author} {\bibinfo {author} {\bibfnamefont {J.}~\bibnamefont
  {Suhonen}},\ }\href {https://doi.org/10.1007/978-3-540-48861-3} {\emph
  {\bibinfo {title} {{From Nucleons to Nucleus}}}},\ Theoretical and
  Mathematical Physics\ (\bibinfo  {publisher} {Springer},\ \bibinfo {address}
  {Berlin, Germany},\ \bibinfo {year} {2007})\BibitemShut {NoStop}%
\bibitem [{\citenamefont {Ring}\ and\ \citenamefont {Schuck}(1980)}]{RiSc80}%
  \BibitemOpen
  \bibfield  {author} {\bibinfo {author} {\bibfnamefont {P.}~\bibnamefont
  {Ring}}\ and\ \bibinfo {author} {\bibfnamefont {P.}~\bibnamefont {Schuck}},\
  }\href {https://www.springer.com/us/book/9783540212065} {\emph {\bibinfo
  {title} {{The Nuclear Many-Body Problem}}}}\ (\bibinfo  {publisher}
  {Springer-Verlag Berlin Heidelberg},\ \bibinfo {year} {1980})\BibitemShut
  {NoStop}%
\bibitem [{\citenamefont {Stedman}(1990)}]{Stedman1990}%
  \BibitemOpen
  \bibfield  {author} {\bibinfo {author} {\bibfnamefont {G.~E.}\ \bibnamefont
  {Stedman}},\ }\href
  {https://www.cambridge.org/us/academic/subjects/physics/theoretical-physics-and-mathematical-physics/diagram-techniques-group-theory}
  {\emph {\bibinfo {title} {{Diagram techniques in group theory}}}}\ (\bibinfo
  {publisher} {Cambridge University Press},\ \bibinfo {year}
  {1990})\BibitemShut {NoStop}%
\bibitem [{\citenamefont {Van~Dyck}\ and\ \citenamefont
  {Fack}(2003)}]{vandyck03}%
  \BibitemOpen
  \bibfield  {author} {\bibinfo {author} {\bibfnamefont {D.}~\bibnamefont
  {Van~Dyck}}\ and\ \bibinfo {author} {\bibfnamefont {V.}~\bibnamefont
  {Fack}},\ }\href {https://doi.org/10.1016/S0010-4655(02)00733-6} {\bibfield
  {journal} {\bibinfo  {journal} {Comp. Phys. Comm.}\ }\textbf {\bibinfo
  {volume} {151}},\ \bibinfo {pages} {354} (\bibinfo {year}
  {2003})}\BibitemShut {NoStop}%
\bibitem [{\citenamefont {Tichai}\ \emph {et~al.}(2019)\citenamefont {Tichai},
  \citenamefont {Ripoche},\ and\ \citenamefont {Duguet}}]{Tichai:2019ksh}%
  \BibitemOpen
  \bibfield  {author} {\bibinfo {author} {\bibfnamefont {A.}~\bibnamefont
  {Tichai}}, \bibinfo {author} {\bibfnamefont {J.}~\bibnamefont {Ripoche}},\
  and\ \bibinfo {author} {\bibfnamefont {T.}~\bibnamefont {Duguet}},\ }\href
  {https://doi.org/10.1140/epja/i2019-12758-6} {\bibfield  {journal} {\bibinfo
  {journal} {Eur. Phys. J. A}\ }\textbf {\bibinfo {volume} {55}},\ \bibinfo
  {pages} {90} (\bibinfo {year} {2019})}\BibitemShut {NoStop}%
\bibitem [{\citenamefont {Parzuchowski}\ \emph
  {et~al.}(2017{\natexlab{b}})\citenamefont {Parzuchowski}, \citenamefont
  {Stroberg}, \citenamefont {Navr{\'{a}}til}, \citenamefont {Hergert},\ and\
  \citenamefont {Bogner}}]{Parzuchowski:2017wcq}%
  \BibitemOpen
  \bibfield  {author} {\bibinfo {author} {\bibfnamefont {N.~M.}\ \bibnamefont
  {Parzuchowski}}, \bibinfo {author} {\bibfnamefont {S.~R.}\ \bibnamefont
  {Stroberg}}, \bibinfo {author} {\bibfnamefont {P.}~\bibnamefont
  {Navr{\'{a}}til}}, \bibinfo {author} {\bibfnamefont {H.}~\bibnamefont
  {Hergert}},\ and\ \bibinfo {author} {\bibfnamefont {S.~K.}\ \bibnamefont
  {Bogner}},\ }\href {https://doi.org/10.1103/PhysRevC.96.034342} {\bibfield
  {journal} {\bibinfo  {journal} {Phys. Rev. C}\ }\textbf {\bibinfo {volume}
  {96}},\ \bibinfo {pages} {034324} (\bibinfo {year}
  {2017}{\natexlab{b}})}\BibitemShut {NoStop}%
\bibitem [{\citenamefont {Hergert}\ \emph
  {et~al.}(2016{\natexlab{b}})\citenamefont {Hergert}, \citenamefont {Bogner},
  \citenamefont {Morris}, \citenamefont {Schwenk},\ and\ \citenamefont
  {Tsukiyama}}]{Hergert:2015awm}%
  \BibitemOpen
  \bibfield  {author} {\bibinfo {author} {\bibfnamefont {H.}~\bibnamefont
  {Hergert}}, \bibinfo {author} {\bibfnamefont {S.~K.}\ \bibnamefont {Bogner}},
  \bibinfo {author} {\bibfnamefont {T.~D.}\ \bibnamefont {Morris}}, \bibinfo
  {author} {\bibfnamefont {A.}~\bibnamefont {Schwenk}},\ and\ \bibinfo {author}
  {\bibfnamefont {K.}~\bibnamefont {Tsukiyama}},\ }\href
  {https://doi.org/10.1016/j.physrep.2015.12.007} {\bibfield  {journal}
  {\bibinfo  {journal} {Phys. Rep.}\ }\textbf {\bibinfo {volume} {621}},\
  \bibinfo {pages} {165} (\bibinfo {year} {2016}{\natexlab{b}})}\BibitemShut
  {NoStop}%
\bibitem [{\citenamefont {Beazley}\ \emph {et~al.}(2018)\citenamefont {Beazley}
  \emph {et~al.}}]{PLY}%
  \BibitemOpen
  \bibfield  {author} {\bibinfo {author} {\bibfnamefont {D.~M.}\ \bibnamefont
  {Beazley}} \emph {et~al.},\ }\href {https://github.com/dabeaz/ply} {\bibinfo
  {title} {{PLY -- Python Lex-Yacc}}} (\bibinfo {year} {2018}),\ \bibinfo
  {note} {\url{https://github.com/dabeaz/ply}}\BibitemShut {NoStop}%
\end{thebibliography}%
\endgroup

\end{document}